\documentclass[manuscript]{acmart}

\usepackage{bm}
\usepackage{nicefrac}
\usepackage{longtable}
\usepackage{array}
\usepackage{bbm}
\usepackage{enumerate}
\usepackage{subcaption}
\usepackage{todonotes}

\newcommand{\cD}{\mathcal{D}}

\newcommand{\cM}{\mathcal{M}}
\newcommand{\cR}{\mathcal{R}}

\newcommand{\RR}{\mathbb{R}}
\newcommand{\ZZ}{\mathbb{Z}}

\newcommand{\schoolsize}{\vert S\vert}
\newcommand{\blocksize}{\vert B\vert}
\newcommand{\groupsize}{\vert G\vert}
\newcommand{\curr}{\text{curr}}

\newcommand{\pr}[1]{\operatorname{Pr}\left(#1\right)}
\newcommand{\ee}[1]{\mathbb{E}\left[#1\right]}
\newcommand{\empe}[1]{\hat{\mathbb{E}}\left[#1\right]}
\newcommand{\lap}[1]{\operatorname{Lap}\left(#1\right)}
\newcommand{\geom}[1]{\operatorname{Geom}\left(#1\right)}
\newcommand{\norm}[1]{\left\lVert#1\right\rVert}

\newcommand{\di}[1]{D\left(#1\right)}

\newcommand{\assign}[1]{\bm{x}^*\left(#1\right)}
\newcommand{\districtcount}[1]{Q_{#1}}
\newcommand{\relu}[1]{\left(#1\right)_+}
\newcommand{\blockcount}[2]{N_{#1,#2}}
\newcommand{\schoolcount}[2]{Q_{#1,#2}}
\newcommand{\neigbor}[2]{A_{#2}\left(#1\right)}

\newcommand{\currassign}{\bm{x}^{(\curr)}}

\AtBeginDocument{%
  }

\setcopyright{acmcopyright}
\copyrightyear{2023}
\acmYear{2023}
\acmDOI{XXXXXXX.XXXXXXX}

\acmConference{Under Review}{2023}{TBD}
\begin{document}

\title{Impacts of Differential Privacy on Fostering more Racially and Ethnically Diverse Elementary Schools}

\author{Keyu Zhu}
\email{kzhu67@gatech.edu}
\orcid{1234-5678-9012}
\affiliation{%
  \institution{Georgia Institute of Technology}
  \streetaddress{North Avenue}
  \city{Atlanta}
  \state{Georgia}
  \country{USA}
  \postcode{30332}
}

\author{Nabeel Gillani}
\affiliation{%
  \institution{Northeastern University}
  \streetaddress{360 Huntington Ave}
  \city{Boston}
  \country{USA}}
\email{n.gillani@northeastern.edu}

\author{Pascal Van Hentenryck}
\affiliation{%
  \institution{Georgia Institute of Technology}
  \streetaddress{North Avenue}
  \city{Atlanta}
  \state{Georgia}
  \country{USA}
  \postcode{30332}
}

% \affiliation{%

% \affiliation{%

% \affiliation{%

% \affiliation{%

% \affiliation{%

\renewcommand{\shortauthors}{Zhu et al.}

\begin{abstract}
  In the face of increasingly severe privacy threats in the era of data and AI, 
  the US Census Bureau has recently adopted differential privacy, 
  the \emph{de facto}
  standard of privacy protection for the 2020 Census release. 
  Enforcing differential 
  privacy involves adding carefully calibrated random noise to sensitive demographic 
  information prior to its release. This change has the potential to impact policy 
  decisions like political redistricting and other high-stakes practices,
  partly because tremendous federal funds and resources are allocated according to 
  datasets (like Census data) released by the US government. One under-explored yet 
  important application of such data is the redrawing of school attendance boundaries 
  to foster less demographically segregated schools. In this study, we ask: how differential privacy might impact diversity-promoting boundaries in 
  terms of resulting levels of segregation, student travel times, and school switching requirements?  Simulating alternative boundaries using differentially-private student counts across 67 Georgia districts, we find that increasing data privacy requirements decreases the extent to which alternative boundaries might reduce segregation and foster more diverse and integrated schools, largely by reducing the number of students who would switch schools under boundary changes.  Impacts on travel times are minimal.  These findings point to a privacy-diversity tradeoff local educational policymakers may face in forthcoming years, particularly as computational methods are increasingly poised to facilitate attendance boundary redrawings in the pursuit of less segregated schools.  
\end{abstract}





\maketitle

\section{Introduction}

In recent years, the concept of differential privacy \cite{dwork2006calibrating}
has gained widespread attention in the field of data 
privacy. Differential privacy is a mathematical framework that prevents the disclosure 
of sensitive information
while still allowing for meaningful analysis. 

One critical 
real-world  application of differential privacy is implemented by the US Census Bureau \cite{abowd20222020}, 
which adopted differential privacy as a privacy-preserving technique for the release of 2020 Census data. 
The census provides crucial data for a variety of purposes, including the allocation of federal funding
\cite{abowd2019economic}, redistricting for political representation \cite{kenny2021use}, and informing research and decision-making in a wide range of fields. The Census Bureau's use of differential privacy is intended to protect the privacy of individuals while still providing accurate population data.

However, recent studies have raised concerns about the potential negative impacts of applying differential
privacy to certain types of data analyses. In particular, several studies 
\cite{DBLP:conf/fat/PujolMKHMM20,zhu2021bias,ijcai2022p559,kenny2021use,steed2022dp}
have found that differential privacy
can introduce bias or unfairness into the outcomes of different high-stakes analyses. 
This is a significant concern, particularly in the context of important decision-making processes that rely on accurate and unbiased data.

While understudied thus far in the differential privacy context, one such decision-making setting is the redrawing of school attendance boundaries---or school catchment areas---for creating less racially and ethnically segregated schools. Segregation by race and class continues to hamper many school districts even today~\cite{gao2022segregation}, and has long been associated with several adverse consequences in K12 educational settings, namely: the perpetuation of achievement gaps~\cite{reardon2018testgaps}, barriers to accessing bridging social capital~\cite{chetty2022socialcapitalII}, and limited opportunities for helping students build understanding and empathy for those who are different from them~\cite{wells2016benefit}.  Generally, the lines determining school district boundaries themselves assume most responsible for existing levels of segregation, but within-district school attendance boundaries can also play a nontrivial role in impeding integration efforts~\cite{fiel2013boundaries, monarrez2021boundaries}.  Even with increasing levels of school choice across the US in recent years~\cite{houlgrave2021choice}, boundary-based assignment continues to be the norm~\cite{nces2021choicefacts}, suggesting attendance boundaries are a powerful ``default setting'', and hence, lever for reducing segregating and advancing more equitable education policies.

Still, changing attendance boundaries can be a highly political and contentious process~\cite{bridges2016eden,zhang2008flight,baltimore2019}.  Yet recent findings using computational boundary simulation models have highlighted the possibility of drawing attendance boundaries in ways that might simultaneously reduce racial/ethnic segregation \textit{and} travel times~\cite{gillani2023redrawing}, sparking interest from school districts to explore how these models might support their desegregation policymaking efforts~\cite{gillani2023las}.  

Of course, to the extent that districts are the primary users of these models, and their uses remain internal within the districts, it is perhaps unclear why differential privacy is relevant here: districts generally have exact, geocoded counts of students by race and ethnicity, and exclusively internal uses obviate the need to obfuscate such counts with noise for privacy-preservation purposes.  Yet districts may lack the technical infrastructure and expertise to adapt and apply redistricting models on their own.  For assistance, they often hire outside consultants, yet even these consultants may use manual procedures and/or lack the necessary tools and expertise.  Researchers and technologists might help bridge these gaps by creating new, widely-available platforms that make it easy for districts to upload their latest geocoded student counts and existing attendance boundaries and obtain simulated, diversity-promoting boundary changes.  Indeed, recent evidence suggests that there may be latent, and perhaps even growing, demand for such tools across many districts~\cite{gillani2023las}.  Even if they are only asked to upload aggregated students counts, school districts may have privacy concerns, especially if certain geographies have very small numbers of students belonging to certain racial groups.  Therefore, equipping such platforms with differential privacy may help increase trust and usage among districts, and ultimately, advance practical school desegregation boundary planning efforts.  

A less forward-looking motivation involves the fact that even school districts with exact geocoded student counts may still turn to Census and American Community Survey (ACS) to approximate demographics they do not have available at the student-level--- like socioeconomic status (SES).  Many districts have turned to SES integration as a primary objective given both the important role it can have on fostering more equitable education and life outcomes~\cite{chetty2022socialcapitalII,yeung2022france} and the controversial legal landscape around using student race/ethnicity in assignment policymaking~\cite{pics_kennedy}.  Districts will sometimes factor in SES, often derived from a composition of several ACS block group-level variables like parental educational attainment, English speaking status, and household income~\cite{quick2016cps}, into student assignment policymaking like boundary planning.  Of course, these publicly-derived datasets may already be released using disclosure avoidance systems, meaning privacy-preservation techniques could at present be influencing student assignment policymaking in ways that are not currently well understood.

These scenarios motivate the need to empirically investigate how imposing differential privacy in school redistricting might affect potential gains (in terms of demographic integration in schools) and costs (in terms of travel times and school switching).  We focus primarily on the first motivating scenario and build on the data and models presented in~\cite{gillani2023redrawing} to explore how differential privacy might affect redistricting efforts to reduce White/non-White segregation across 67 school districts in one state: Georgia.  Georgia has a storied history with both civil rights and school desegregation~\cite{ga2007desegregation}, and continues to be an important state in the movement for more racial justice and equity~\cite{arbery2022nyt}.  Our specific question is: how differential privacy might impact diversity-promoting boundaries in 
  terms of resulting levels of segregation, student travel times, and school switching requirements?

Simulations of alternative boundaries across these Georgia school districts reveal that imposing differential privacy generally decreases the extent to which alternative boundaries foster greater racial/ethnic integration---seemingly by reducing the number of students who are redistricted to different schools---while having minimal effects on expected travel times.  We find similar results when optimizing for SES integration, which we present in the Appendix.  Together, these findings point to a privacy-diversity tradeoff that districts and researchers may face in the coming years if differential privacy is used in computationally-supported desegregation policymaking efforts.

% Missing figures and tables are deferred to the appendix.

\section{Related Work}
\label{sec:related}

\subsection{Impacts of differential privacy in high-stakes decision-making}
Motivated by widespread applications of differential privacy made by 
statistical agencies \cite{abowd20222020} 
and corporations \cite{appledp}, there has been a stream of research about
evaluating the impacts of differential privacy on downstream tasks. 
\cite{DBLP:conf/fat/PujolMKHMM20} presents an empirical study 
of multiple decision-making tasks, e.g., minority language voting rights and fund allocation, 
based on privatized data and demonstrates that disparities in outcomes may
arise due to adoption of differential privacy.
For example, \cite{kenny2021use} systematically analyzes how 
the U.S. Census Bureau’s latest Disclosure Avoidance System (DAS) \cite{abowd20222020} 
powered by differential privacy
affects
the redrawing of electoral districts.  The authors identify how DAS might prevent policymakers from 
creating districts of equal population---and hence, potentially violate the ``One Person, One Vote" standard for political equality.
Following~\cite{DBLP:conf/fat/PujolMKHMM20}'s illumination of how
differential privacy may exert disparate impacts, 
\cite{steed2022dp} conducts an in-depth analysis of 
Title I fund allocation (over \$11.6 billion in 2021) under privacy protection, surfacing how applications of differential privacy
might lead to the misallocation of federal education funds 
due to existing data deviations. Building upon these critical findings, recent studies such as
\cite{ijcai2021p78,ijcai2022p559} focus on 
designing proper mechanisms to
remedy the negative effects brought by differential privacy, especially 
unfairness in fund allocation.

\subsection{Algorithmic school redistricting}
This paper aims to investigate a demographic-related downstream task, i.e.,
redrawing of elementary school attendance boundaries for 
creating more racial and ethnic diversity, and analyze the potential
influences of differential privacy on its performance. 
In terms of this specific task, this study builds upon the data and models from~\cite{gillani2023redrawing}, which takes an initial step
towards simulating alternative attendance boundaries
optimized to achieve racial and ethnic desegregation across multiple US school
districts.  Our present study is relevant in light of this prior work, and the follow-on interest from districts to explore how algorithmic redistricting models might help reveal new pathways to desegregation policymaking in their own contexts~\cite{gillani2023las}.  Given this interest, along with ongoing researcher-practitioner partnerships that involve the development and application of computational models for fostering more diverse schools (e.g.,~\cite{allman2022}), in the forthcoming years, there may be demand for tools and systems that flexibly enable districts to directly explore possible diversity-promoting boundary configurations on their own specific datasets.  Privacy preservation (e.g., through the use of differential privacy) is likely to play an important role in fostering the trust needed for districts to feel comfortable using such systems.  Therefore, we seek to fill an important gap in the literature: empirically investigating how differential privacy might affect the outputs of diversity-promoting school redistricting algorithms.

% \todo[color=blue!25]{Add more related works here.}

\section{Preliminaries of Differential Privacy}
\label{sec:preliminary}

Differential privacy is a concept in data privacy that aims to protect the privacy 
of individuals when their data is being used for analysis. It ensures that 
statistical queries made on a dataset do not reveal too
much sensitive information about any individual. The mathematical definition
of differential privacy is formally presented as follows.
\begin{definition}[Differential Privacy \cite{dwork2006calibrating}]
  A randomized mechanism $\cM : \cD \mapsto \cR$ with domain $\cD$ and range 
  $\cR$ is $\epsilon$-differentially private, if, for any two 
  neighboring datasets
  $D \sim D'$, which differ in only one record, 
  and any possible outcome $R \subseteq \cR$:
  \begin{equation*}
    \pr{\cM(D) \in R} \leq  \exp\left(\epsilon\right)\cdot
      \pr{\cM(D') \in R}\,,
  \end{equation*}
  where the probabilities above are taken over the randomness of the mechanism $\cM$.
\end{definition}
The key parameter $\epsilon$, often known as privacy budget, is a
non-negative real number, which 
quantifies the privacy loss of the randomized algorithm $\cM$.
It measures the strength of privacy protection provided by the algorithm, with
smaller values of $\epsilon$ indicating stronger privacy guarantees.

The Laplace mechanism, which employs carefully calibrated Laplace noise
to obfuscate query answers,
is one of the most commonly-used differentially private mechanisms both in 
theory and practice.

\begin{theorem}[Laplace Mechanism \cite{dwork2006calibrating}]
	\label{th:m_lap} 
	Let $f: \cD \mapsto \RR^n$ be a numeric query. The Laplace mechanism
 achieves $\epsilon$-differential privacy with its output $f(D) + \bm{\eta}$,
 where, for any $i\in[n]$, the entry $\eta_i$ of $\bm{\eta} \in \RR^n$ is an i.i.d. 
 sample drawn from the Laplace distribution $\lap{\nicefrac{\Delta_f}{\epsilon}}$.
\end{theorem}

In the definition above, 
the parameter $\Delta_f$, often called (global) sensitivity,
measures the maximum amount that the query answer can change 
due to addition or removal of
an individual record in the dataset, i.e.,
\begin{equation*}
    \Delta_f\coloneqq \max_{D\sim D'}~\norm{f(D)-f(D')}\,.
\end{equation*}

This work considers a discrete variant of the Laplace mechanism, known as 
the geometric
mechanism, which is capable of preserving both the integrality of its output and 
providing privacy guarantees.

\begin{theorem}[Geometric Mechanism \cite{ghosh2009universally}]\label{thm:geometric_mech}
    Let $f: \cD \mapsto \RR^n$ be a numeric query. The geometric mechanism
    achieves $\epsilon$-differential privacy with its output $f(D) + \bm{\eta}$,
    where, for any $i\in[n]$, the entry $\eta_i$ of $\bm{\eta} \in \RR^n$ is an 
    i.i.d. sample drawn from the two-sided geometric distribution 
    $\geom{\alpha}$ with $\alpha=\exp\left(\nicefrac{\epsilon}{\Delta_f}\right)$.
    The probability mass function of a random variable $Y\sim\geom{\alpha}$
    is shown as follows
    \begin{equation*}
        \pr{Y=k}=\left(\frac{\alpha-1}{\alpha+1}\right)\alpha^{-\vert k\vert}\,,
        \qquad\forall~k\in \ZZ\,.
    \end{equation*}
\end{theorem}

Besides, this work is dependent on a crucial property of differential privacy,
which is its immunity to post-processing. 
This desirable property states that a differentially private output can be
arbitrarily transformed, using some data-
independent function, without affecting its privacy guarantees.

\begin{theorem}[Post-processing Immunity \cite{dwork2006calibrating}]\label{thm:post_processing}
    Let $\cM: \cD \mapsto\cR$ be a randomized mechanism that is $\epsilon$-differentially private. Let $g : \cR \mapsto\cR'$ 
    be an arbitrary mapping. Then, $g \circ \cM : \cD \mapsto\cR'$
    is $\epsilon$-differentially private.
\end{theorem}

\section{Problem Settings}
\label{sec:problem}
This paper uses the following notation: boldface symbols denote vectors or
matrices, while italic symbols are used to denote scalars or random variables. 
This work focuses on simulating alternative elementary school attendance boundaries within a collection of school districts. Formalizing the model presented in~\cite{gillani2023redrawing}, let 
$B$ and $S$ stand for the sets of Census blocks and elementary schools,
both of which belong to a particular school district,
respectively. A valid school assignment can be characterized by a 
$\blocksize\times\schoolsize$ binary matrix, i.e., $\bm{x}\in \{0,1\}^{\blocksize
\times\schoolsize}$, with each row summed up to $1$. The entry $x_{b,s}$ equals $1$,
if block $b$ is assigned to school $s$, or $0$ otherwise. For convenience, let
$\currassign$ represent the current school assignment. In addition, $G$ 
is to used to represent a set of races and ethnicities, including White, Black, Asian, 
Native and Hispanic/Latinx.

In pursuit of more racially and ethnically diverse elementary schools,
\cite{gillani2023redrawing} explores alternative attendance boundaries different from the current ones and 
proposes a novel optimization approach depicted in
Equation \eqref{eq:rezoning_opt}, which takes into consideration multiple realistic constraints.
This optimization problem aims to minimize White/non-White segregation, represented by the dissimilarity index across both groups \cite{duncan1955methodological}, i.e.,
\begin{equation*}
    \di{\bm{x}}\coloneqq\frac{1}{2}\sum_{s\in S}
      \left \vert \frac{\schoolcount{w}{s}}{\districtcount{w}}-
      \frac{\schoolcount{nw}{s}}{\districtcount{nw}}\right \vert\,,
\end{equation*}
where, given a school assignment $\bm{x}$, 
the quantities, $\schoolcount{w}{s}$ and $\schoolcount{nw}{s}$ defined in Equations
\eqref{eq:wht_student} -- \eqref{eq:non_wht_student},
stand for the numbers of White and non-White students attending school $s$ respectively.
Likewise, $\districtcount{w}$ and $\districtcount{nw}$ given by Equation 
\eqref{eq:total_student} indicate the total numbers of
White and non-White students across the school district.
    Equations \eqref{eq:one_choice} -- \eqref{eq:contiguity} present
a list of the constraints that a feasible school assignment is expected to meet:
\begin{enumerate}
    \item[\eqref{eq:one_choice}] {\bf One school per block.} Each Census block should be assigned to only one elementary school.
    \item[\eqref{eq:travel_time}] {\bf Maximum travel time increases.} A reasonable
    school assignment is not supposed to increase estimated travel
    times by more than $\alpha_t$ for any given family living in the
    school district. For any $b\in B$ and $s\in S$, 
    $t_{b,s}$ represents the estimated travel time from
    block $b$ to school $s$.
    \item[\eqref{eq:school_size}] {\bf Maximum school size increases.} 
    To ensure the quality of education, the increase of the total population at a given school 
    resulting from re-assignments of blocks should not exceed 
    $\alpha_P$ of its current population. The 
    total population of a school is the
    combined student counts of all census blocks
    that have been assigned to that school.
    Besides, $\blockcount{T}{b}$ represents the total count of students living in block $b$.
    \item[\eqref{eq:contiguity}] {\bf Contiguity.} If block $b$
    is assigned to school $s$, there should be at least one neighbor of
    $b$ in the induced graph of blocks, which is closer\footnote{The distance between
     block $b$ and school $s$ is measure by the shortest path connecting $b$ to the ``root" block where $s$ is located.}
    to and assigned to 
    school $s$ as well. The set $\neigbor{b}{s} $ is a collection of
    the neighbors of block $b$, which are closer to school $s$.
\end{enumerate}

\begin{subequations}\label{eq:rezoning_opt}
    \begin{align}
      \assign{\bm{N}}=\underset{\bm{x}}{\arg\min}&~~\frac{1}{2}\sum_{s\in S}
      \left \vert \frac{\schoolcount{w}{s}}{\districtcount{w}}-
      \frac{\schoolcount{nw}{s}}{\districtcount{nw}}\right \vert \\
      \text{s.t.}&~~\sum_{s\in S}x_{b,s} = 1\,, & \forall~b\in B\,,\label{eq:one_choice}\\
      &~~\sum_{s\in S}x_{b,s}\cdot t_{b,s}\leq \left(1+\alpha_t\right)
      \sum_{s\in S}x_{b,s}^{(\curr)}\cdot t_{b,s}\,, & \forall~b\in B\,,\label{eq:travel_time}\\
      &~~\sum_{b\in B} 
      x_{b,s}\cdot \blockcount{T}{b}\leq \left(1+\alpha_P\right)
      \sum_{b\in B}  x_{b,s}^{(\curr)} \cdot\blockcount{T}{b}\,, & 
      \forall~s\in S\,,\label{eq:school_size}\\
      &~~x_{b,s}\leq \sum_{b'\in \neigbor{b}{s}} x_{b',s}\,, & \forall~b\in B, s\in S\,,
      \label{eq:contiguity}\\
       &\schoolcount{w}{s} = \sum_{b\in B}x_{b,s}\cdot \blockcount{w}{b}\,,
      &\forall~s\in S\,,\label{eq:wht_student}\\
      &\schoolcount{nw}{s}=\sum_{b\in B}x_{b,s}\cdot\left(\blockcount{T}{b}- 
      \blockcount{w}{b}\right)\,,
      &\forall~s\in S\,,\label{eq:non_wht_student}\\
      &\districtcount{g}=\sum_{s\in S}\schoolcount{g}{s}\,, &\forall~g\in\left\{w, nw\right\}\,,
      \label{eq:total_student}\\
      &~~x_{b,s} \in \left\{0,1\right\}\,, &\forall~b\in B, s\in S\,.
    \end{align}
\end{subequations}

The solution $\bm{x}^*(\bm{N})$ 
to Equation \eqref{eq:rezoning_opt} is referred to as the non-private school assignment,
in order to distinguish it from the differentially private 
one that will be discussed later. 

\begin{figure}
    \centering
    \includegraphics[width=.7\linewidth]{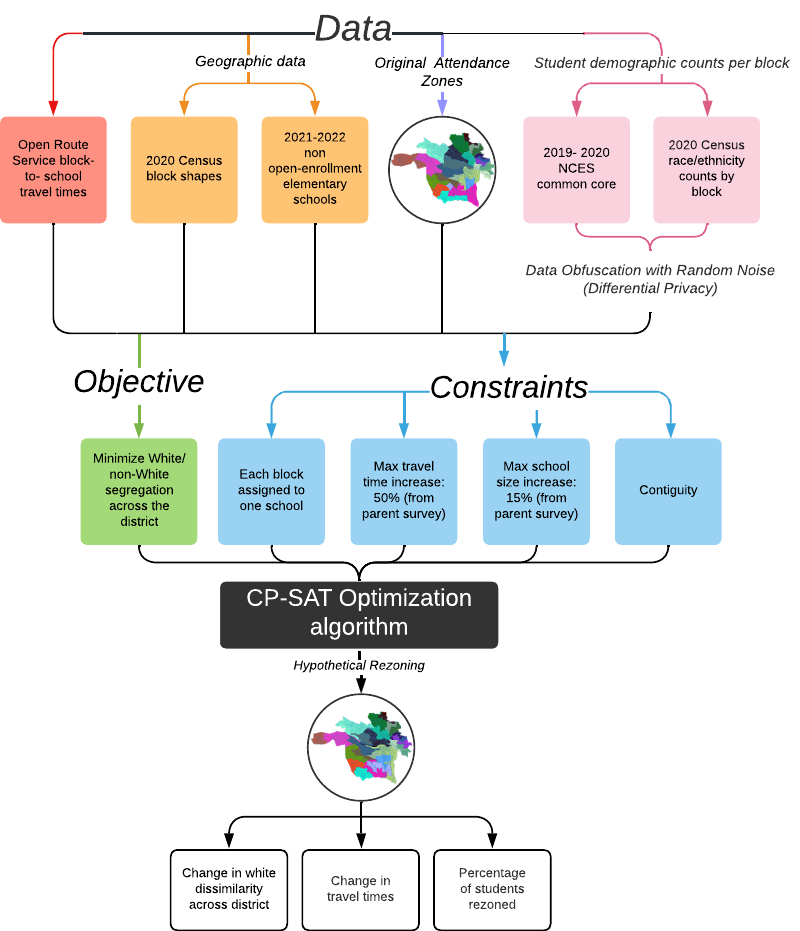}
    \caption{Input data, objective function, constraints, and outcome measures from the optimization
    model in Equation \eqref{eq:rezoning_opt}. 
    Note that, in order to retrieve private school assignments, the student counts
    should be perturbed by random noise
    prior to optimization.  Figure adapted from~\cite{gillani2023redrawing}.}
    \label{fig:flowchart}
\end{figure}

Notice that, to emphasize its dependence
on the student counts $\bm{N}$, the non-private school assignment is expressed as
a function of $\bm{N}$, which is essentially a $\left(\groupsize+1\right)
\times\blocksize$
matrix with each column $\bm{N}_{\cdot,b}$ a record of the students of different races
and ethnicities in $G$, along with the total counts of students living 
in block $b$, i.e.,
\begin{equation*}
    \bm{N}_{\cdot,b}^\top = 
    \left[
    \begin{matrix}
        \bm{N}_{G,b}^\top  & \blockcount{T}{b}
    \end{matrix}
    \right]\,,\qquad\forall~b\in B\,.
\end{equation*}
In the context of data privacy, the demographics in the form of
$\bm{N}$ can reveal sensitive information about individuals such as 
race/ethnicity and thus require privacy protection prior to data release. 
This work assumes that, in the context of democratizing access to algorithmic redistricting tools for promoting more diverse schools, school districts may wish to share student counts $\bm{N}$ in a differentially private manner, which could be accomplished by designing and implementing a platform that applies the geometric mechanism in Theorem \ref{thm:geometric_mech} to district-uploaded data and uses this as the basis for simulating alternative boundaries.
Specifically, the student counts $\bm{N}$, as a query answer $f(D)$, can preserve
$\epsilon$-differential privacy, 
if i.i.d. two-sided geometric noise $\eta\sim\geom{\exp\left(\nicefrac{\epsilon}{\Delta_f}\right)}$
with $\Delta_f=2$
is added to each of their entries. The reason why the sensitivity associated with $\bm{N}$ 
equals $2$ is that addition or removal of an individual record from $D$, say a student who
lives in block $b$ and belongs to racial group $g$, would affect exactly two entries of $\bm{N}$, 
$\blockcount{g}{b}$ and $\blockcount{T}{b}$. Moreover, a simple post-processing step
\footnote{$\relu{\cdot}$ is an entry-wise operation of taking the positive part of the input.}
should be performed ahead of data release to restore non-negativity, i.e., 
$\tilde{\bm{N}}=\relu{\bm{N}+\bm{\eta}}$, such that there is no negative count in $\tilde{N}$.
By replacing $\bm{N}$ with their private
counterparts $\tilde{\bm{N}}$, the optimized school assignment $\assign{\tilde{\bm{N}}}$
ensures $\epsilon$-differential privacy due to post-processing immunity\footnote{
For $\assign{\tilde{\bm{N}}}$, the optimization problem in 
Equation
\eqref{eq:rezoning_opt} can be considered as a specific form of post-processing
applied to $\tilde{N}$.} in Theorem 
\ref{thm:post_processing} and is thus referred to as the private school assignment.
Figure \ref{fig:flowchart}, adapted from~\cite{gillani2023redrawing}, provides a graphic illustration of the systematic
procedure for generating private school assignments as inputs into the downstream rezoning process described above.

Therefore, we end up with three different school assignment scenarios:
\begin{itemize}
    \item {\bf current school assignment} $\currassign$,
    \item {\bf non-private school assignment} $\assign{\bm{N}}$,
    \item {\bf private school assignment} $\assign{\tilde{\bm{N}}}$.
\end{itemize}
The main objective of this work is to evaluate the potential impacts of differential privacy
on redrawing the school attendance boundaries by comparing private and non-private school
assignments in different aspects, including but not limited to racial and ethnic diversity.
On many occasions, the current school assignment serves as the benchmark against the other two
school assignments.

\section{Experiments and Results}
\label{sec:experiment}
 This paper 
 focuses on 968 elementary schools across 67 school districts in Georgia, using the datasets presented 
in \cite{gillani2023redrawing}: 
2021-2022 elementary school attendance boundaries, 2019-2020 publicly-available
school attendance data by race/ethnicity (White, Black, Asian, Native, and
Hispanic/Latinx), and 2020 Census-level estimates of the number of non-adults per demographic group 
living in each Census block. These datasets are used to estimate the actual populations of students
living in 134,324 Census blocks.  While these are estimates derived from the aforementioned datasets, for the purposes of this study, we refer to them as ``ground truth'' and use them as the basis for imposing differential privacy. Figure \ref{fig:demo_info} presents an overview of the racial
and ethnic identities of the students.
This work utilizes those student counts to derive non-private school assignments across 67
school districts, by
solving the optimization problem in Equation \eqref{eq:rezoning_opt} using the 
CP-SAT constraint programming solver \cite{ortools}. Likewise, private
school assignments can be generated by similar procedures, as depicted in 
the previous section, but based on privatized student counts. It
is noteworthy that, whatever the school assignments, the dissimilarity
indices are evaluated on the ground-truth student counts in the following sections,
rather than the differentially private counts used to create private school
assignments.

In addition, the scenarios of private school assignments
under different levels of privacy protection 
are showcased using two different privacy budgets, namely $\epsilon\in \{2,4\}$,
where the variances of the two-sided geometric noise $\eta$ used for data obfuscation
are around $1.74$ and $0.36$ respectively.
The private school assignment with the privacy budget $\epsilon=2$ provides 
stronger privacy guarantees than the other ($\epsilon=4$).
Due to 
random nature of private school assignments, 200 independent rezonings are 
simulated for each district and privacy budget $\epsilon$. To conclude the experimental
settings, this work permits up to $50\%$ increase in travel time and $15\%$ increase
in school size
due to re-assignments of Census blocks, i.e., $\alpha_t = 50\%$ and $\alpha_P=15\%$, motivated by data obtained from a parent survey in~\cite{gillani2023redrawing} highlighting these as potentially realistic values for such parameters.

\subsection{District-level Results}
\begin{figure}
    \centering
    \includegraphics[width=.7\linewidth]{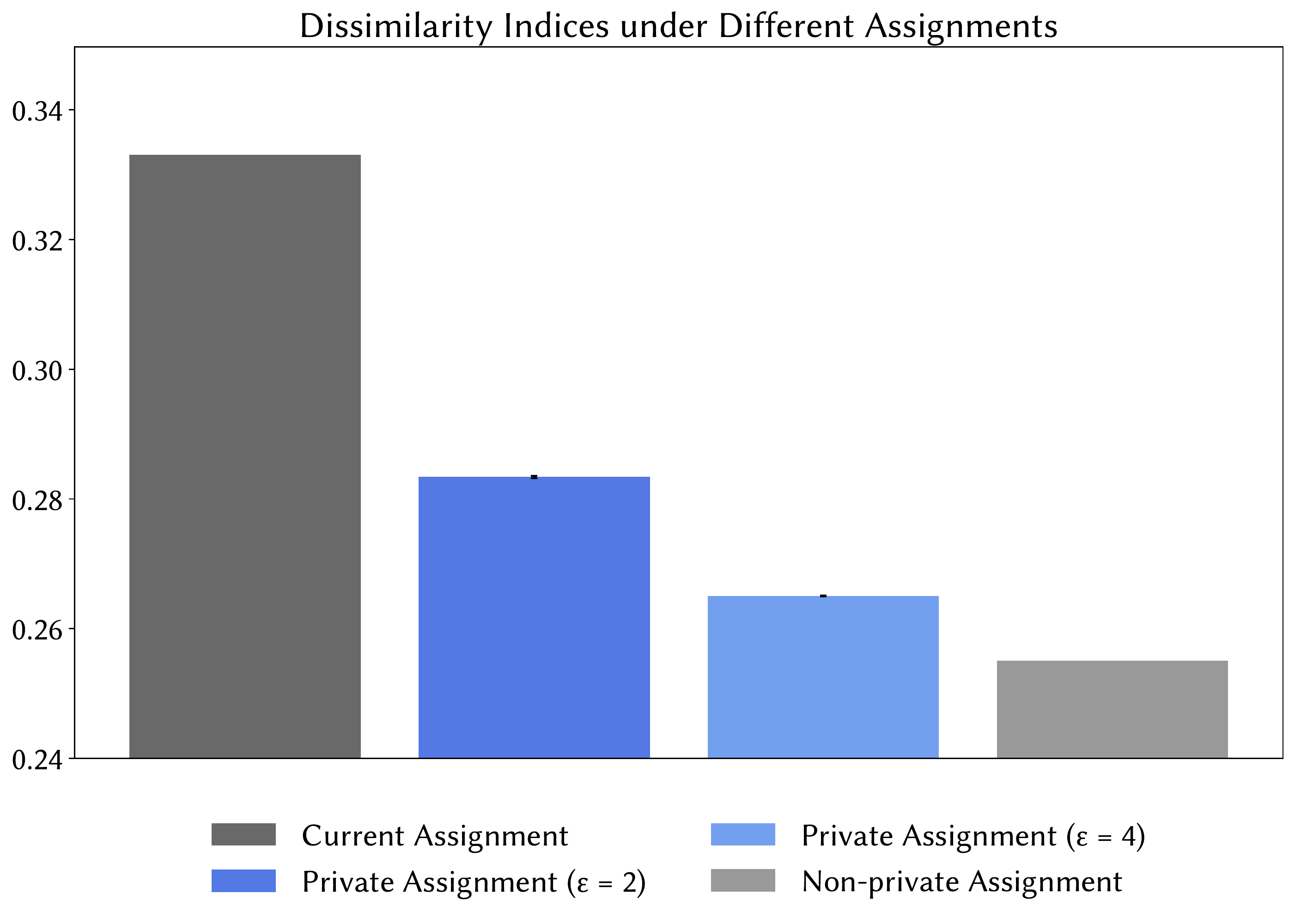}
    \caption{Average White/non-White dissimilarity indices associated with current, non-private, and private 
    school assignments
    with  privacy budgets $\epsilon=2$ and $4$. Error bars depict 95\% bootstrapped
    confidence intervals for results across independent rezonings with random noise added.  The bars represent averages across districts; for a sense of variability by district, please see the extended Table~\ref{tab:long_dissim_idx_info}.}
    \label{fig:dissim_idx}
\end{figure}

For the convenience of presentation, 
this section mainly covers the results associated with the stronger privacy budget $\epsilon=2$,
unless otherwise specified. The primary goal here is to examine the 
impacts of differential privacy on the redrawing of school attendance boundaries
in terms of racial and ethnic desegregation.
Table \ref{tab:long_dissim_idx_info} contains the dissimilarity indices associated with 
the three school assignments of different types across 67 school districts.
In most of the school districts studied in this work,
the performance of the private school assignments is moderate 
and lies somewhere between the non-private and current ones.
To be more accurate, as illustrated in Figure \ref{fig:dissim_idx}, 
the private school assignment would result in more racially 
segregated attendance boundaries across the studied districts than the non-private one: a median
decrease of 14.91\%, versus a median 23.41\% reduction under non-private school assignment. This 
may be due in part to the fact that
the private school assignment assigns 
around 4.75\% fewer Census blocks, on average, to different schools than the 
non-private one.

\begin{figure}
    \centering
    \includegraphics[width=.7\linewidth]{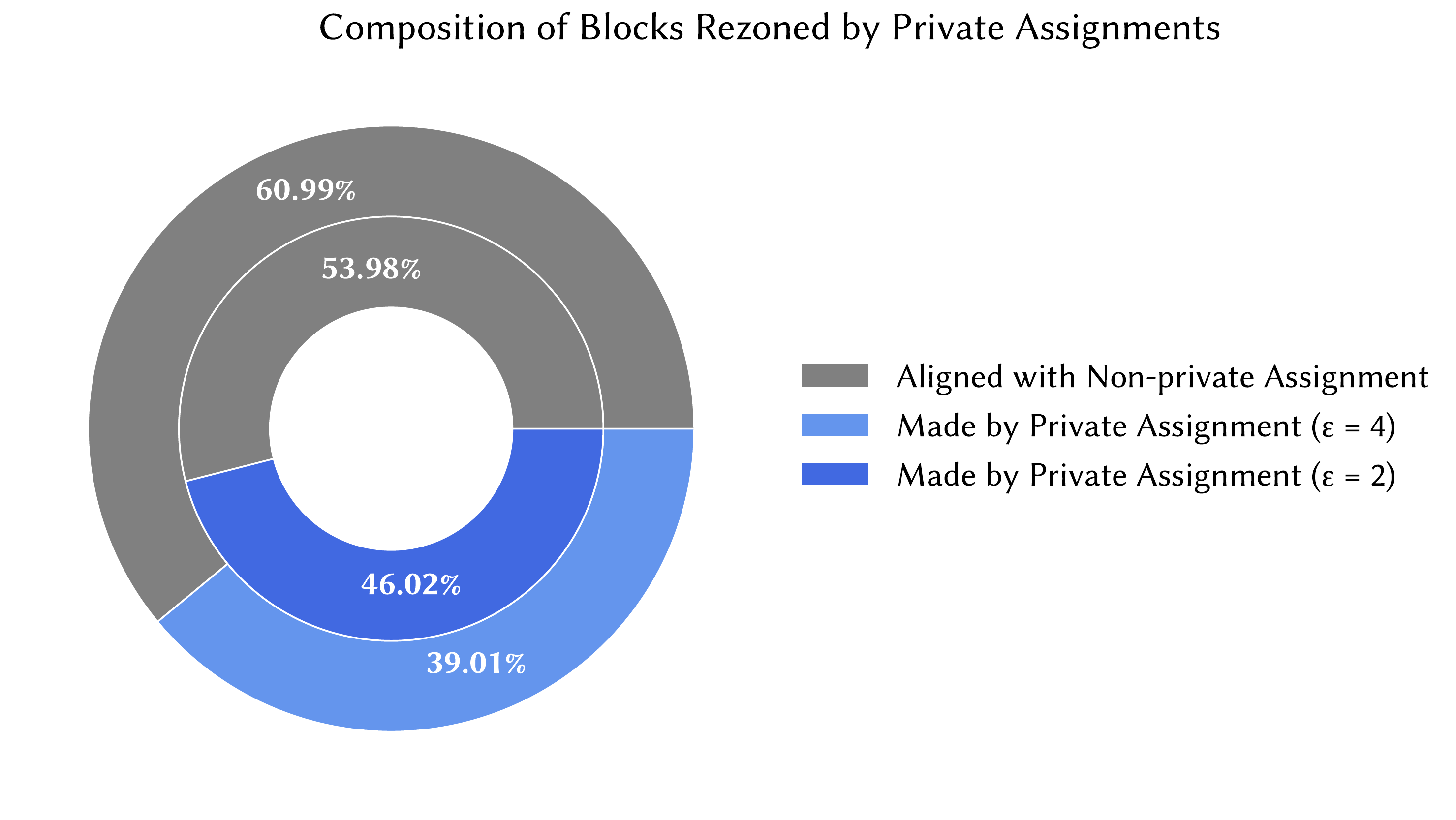}
    \caption{Composition of Census blocks rezoned by private 
        school assignments.
        The inner and outer circles correspond to the private
        school assignments with $\epsilon=2$ and $4$, respectively.}
    \label{fig:block_comp}
\end{figure}

Moreover, notice that the private school assignment
with the privacy budget $\epsilon=4$ outperforms that with $\epsilon=2$
and achieves a smaller performance gap against the non-private school assignment.
This observation can be partially justified by the fact that,
perturbed by less random noise, the private school assignment under
$\epsilon=4$ rezones 732 more Census blocks in expectation and, more importantly,
the re-assignments it makes turn out to be more efficient.
Figure \ref{fig:block_comp} shows that over 60\% of the re-assignments of Census blocks
made by the private school assignment with $\epsilon=4$
coincide with those made by the non-private one while 
that kind of re-assignments only accounts for 
just around 54\% of the total re-assignments 
made by the private one with $\epsilon=2$.
Another intuitive way to understand this observation is that 
the non-private school assignment can somehow be interpreted as a private one,
whose privacy budget equals $+\infty$ because it is derived by a deterministic
mechanism applied to the student counts with no noise added. 
Therefore, as the privacy
budget $\epsilon$ increases
and privacy protection gets weaker, we might expected that the private school
assignment converges to the non-private one.

\begin{figure}
    \centering
    \includegraphics[width=.7\linewidth]{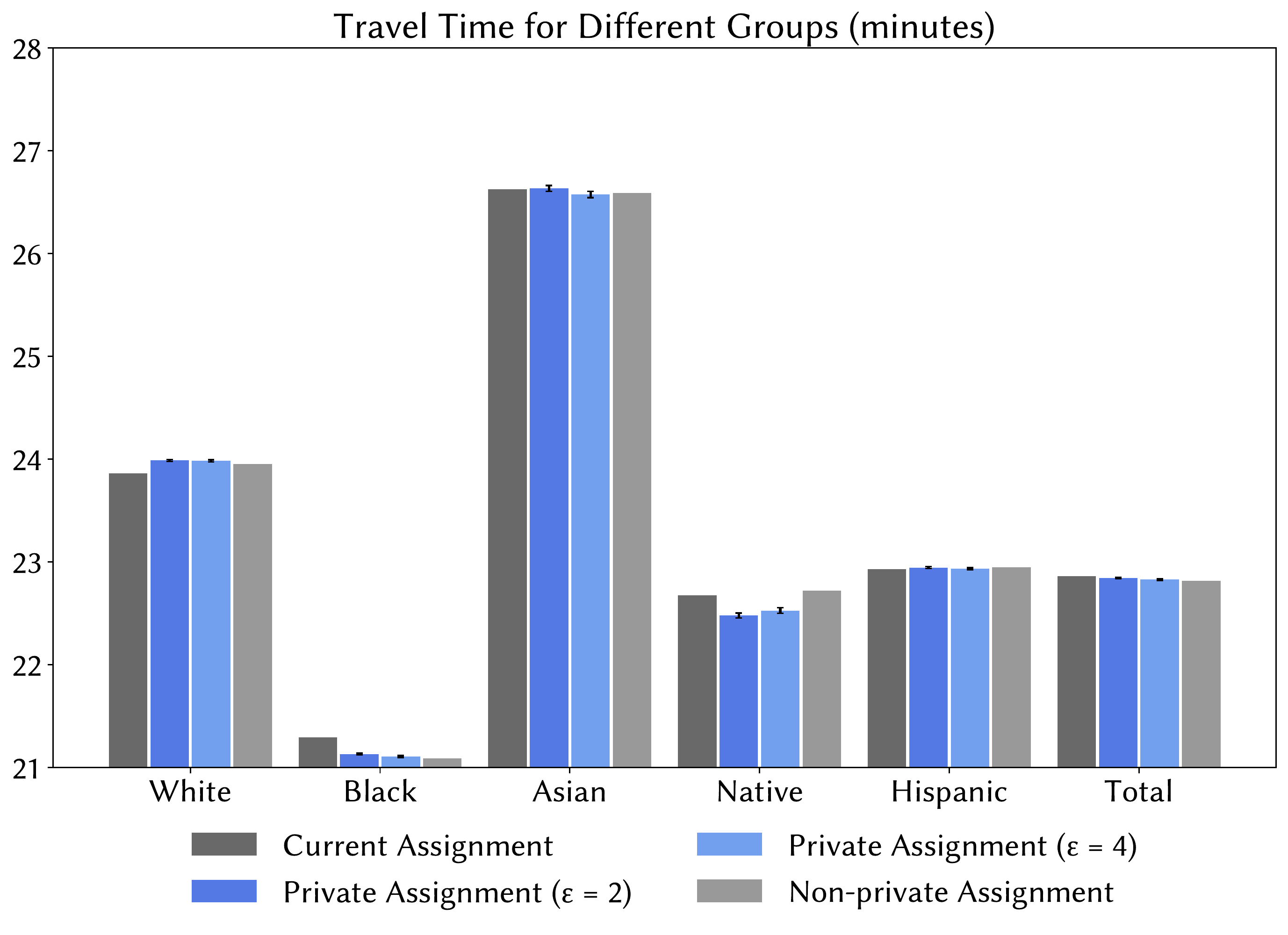}
    \caption{Travel time for different groups. Error bars depict 95\% bootstrapped confidence
    intervals across across independent privacy runs.}
    \label{fig:travel_time}
\end{figure}

\begin{figure}
    \centering
    \includegraphics[width=.7\linewidth]{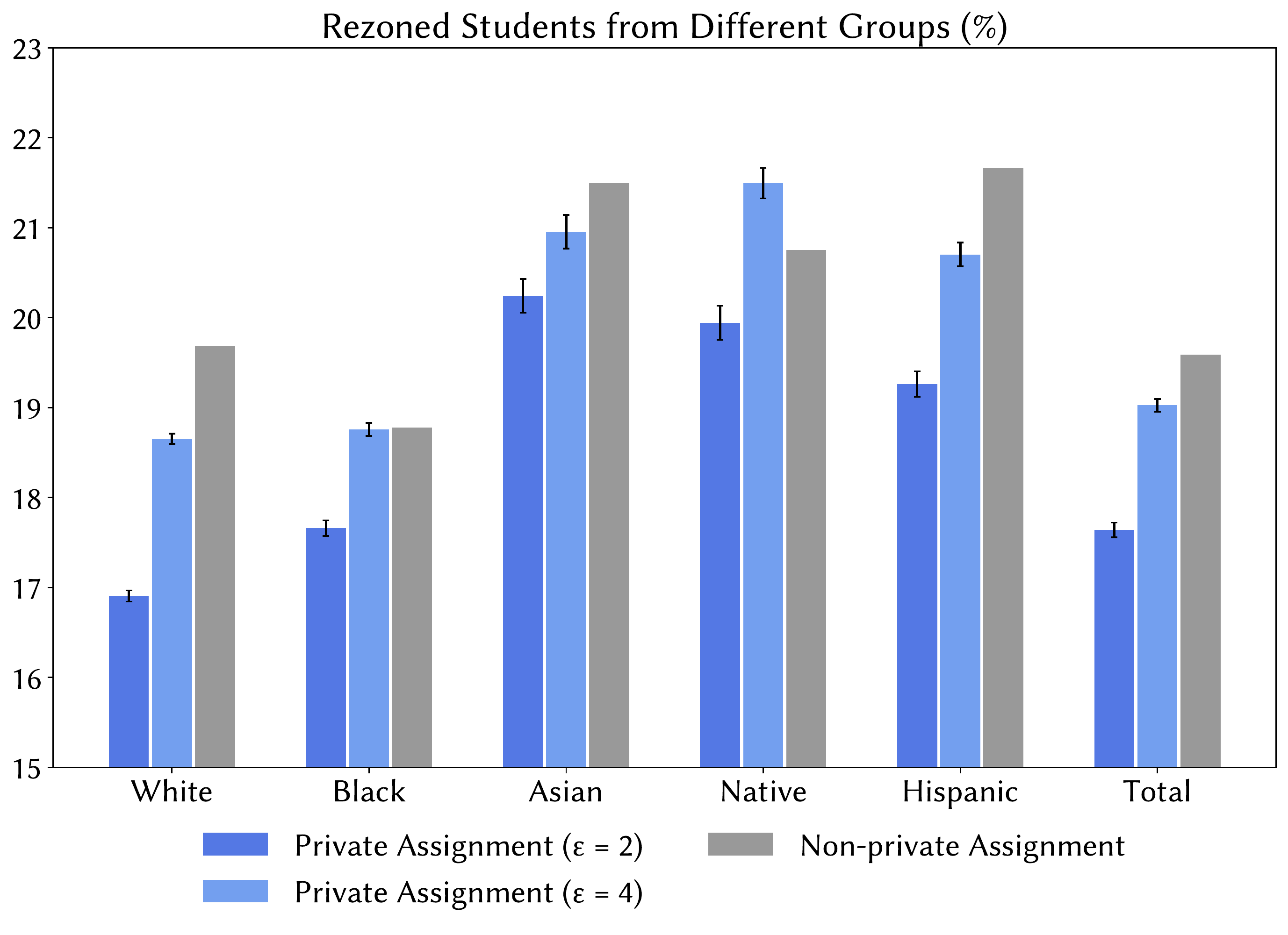}
    \caption{Percentage of the students rezoned across different groups. 
    Error bars depict 95\% bootstrapped confidence intervals across independent privacy runs.}
    \label{fig:rezoned}
\end{figure}

Figure \ref{fig:travel_time} visualizes the changes in travel time resulting from different 
school assignments. Interestingly, both private and non-private boundary changes
do not substantially change travel times, and in many cases, reduce them across demographic 
groups, similar to the findings from~\cite{gillani2023redrawing}. Figure \ref{fig:rezoned} shows that 
around 2\%
fewer students would be re-assigned to other schools by the private school assignment, compared 
with the non-private one,
and the reduction is similar across different demographic groups. An exception occurs under 
the private school assignment with $\epsilon=4$, where more Native American students are affected,
likely due to their small population size.

Next, we hone into specific configurations across districts—namely, the scenario in each district 
that would produce the smallest reduction in segregation due to introduction of differential 
privacy, i.e., the largest difference in dissimilarity index between the private and non-private 
school assignments $\empe{\di{\assign{\tilde{\bm{N}}}}} - \di{\assign{\bm{N}}}$. 
We are interested in better understanding which district-level features are predictive of this difference, if any.  This is important, as it may help shed light on which types of districts the introduction of differential privacy is more or less likely to impact, and how. 

We use linear regression to study which school district features are associated with this difference. 
The selected features can be grouped into the following four categories:
\begin{enumerate}[1)]
    \item the demographic composition of students,
    \item the size of the school district,
    \item the location of the school district,
    \item the current level of racial and ethnic segregation.
\end{enumerate}

% Table generated by Excel2LaTeX from sheet 'res_summary'
\begin{table}[htbp]
  \centering
  \caption{Regression result. $\di{\currassign}$ is a shorthand for the dissimilarity 
  index associated with the current school assignment and the last two columns represent the lower and upper bounds of the 95\% confidence intervals.}
    \begin{tabular}{|l|r|r|>{\raggedleft\arraybackslash}p{1cm}|r|>{\centering\arraybackslash}p{1.6cm}|>{\centering\arraybackslash}p{1.6cm}|}
    \hline
         \textbf{Features} & \textbf{Coefficients} & \textbf{Standard Errors} 
         & $t$ & $p$\textbf{-values} & \multicolumn{2}{c|}{\textbf{95\% Confidence Intervals}} \\
    \hline
    \% White    & 0.0098 & 0.012 & 0.785 & 0.436 &
    -0.015 & 0.035 \\
    \% Black    & 0.0077 & 0.012 & 0.652 & 0.517 & -0.016 & 0.031 \\
    \% Asian    & -3.79E-5 & 0.002 & -0.017 & 0.986 & -0.004 & 0.004 \\
    \% Native   & 0.001 & 0.001 & 0.999 & 0.322 & -0.001 & 0.003 \\
    \% Hispanic    & 0.0034 & 0.007 & 0.498 & 0.621 & -0.01 & 0.017 \\
    \hline
    \# Students    & 0.0026 & 0.004 & 0.729 & 0.469 & -0.005 & 0.01 \\
    \# Elem. Schools
    & -0.0017 & 0.004 & -0.462 & 0.646 & -0.009 & 0.006 \\
    \hline
    $\mathbbm{1}\left\{\text{rural}\right\}$
    & 0.0004 & 0.001 & 0.693 & 0.491 & -0.001 & 0.002 \\
    $\mathbbm{1}\left\{\text{small city}\right\}$
    & -0.0005 & 0     & -0.935 & 0.354 & -0.001 & 0.001 \\
    $\mathbbm{1}\left\{\text{suburban}\right\}$
    & 2.46E-5 & 0.001 & 0.043 & 0.966 & -0.001 & 0.001 \\
    $\mathbbm{1}\left\{\text{urban}\right\}$   & 0.0004 & 0.001 & 0.416 & 0.679 & -0.001 & 0.002 \\
    \hline
    $\di{\currassign}$   & 0.0046 & 0.001 & 4.667 & 0     & 0.003 & 0.007 \\
    \hline
    \end{tabular}%
  \label{tab:regression}%
\end{table}%

Results are summarized in 
Table \ref{tab:regression}. We find that out of the included variables, only the district's 
baseline White/non-White dissimilarity index 
$\di{\currassign}$ is significantly predictive of this 
difference (at $\alpha=0.05$): the higher the $\di{\currassign}$, the less diverse the private school assignment is (i.e., the larger the gap between resulting dissimilarity between the private and non-private 
school assignments).  The 
adjusted $R^2$ of the model is also generally small at approximately 0.32.  Together, these values suggest that the impact of differential privacy on 
potential school integration is likely to vary widely across districts, and will be influenced by 
district-level idiosyncrasies like current boundaries, demographic distributions across 
constituent neighborhoods, and other factors not currently included in our model.

\subsection{Block-level Results}
\begin{figure}
    \centering
    \begin{subfigure}{0.35\linewidth}
        \includegraphics[width=\linewidth]{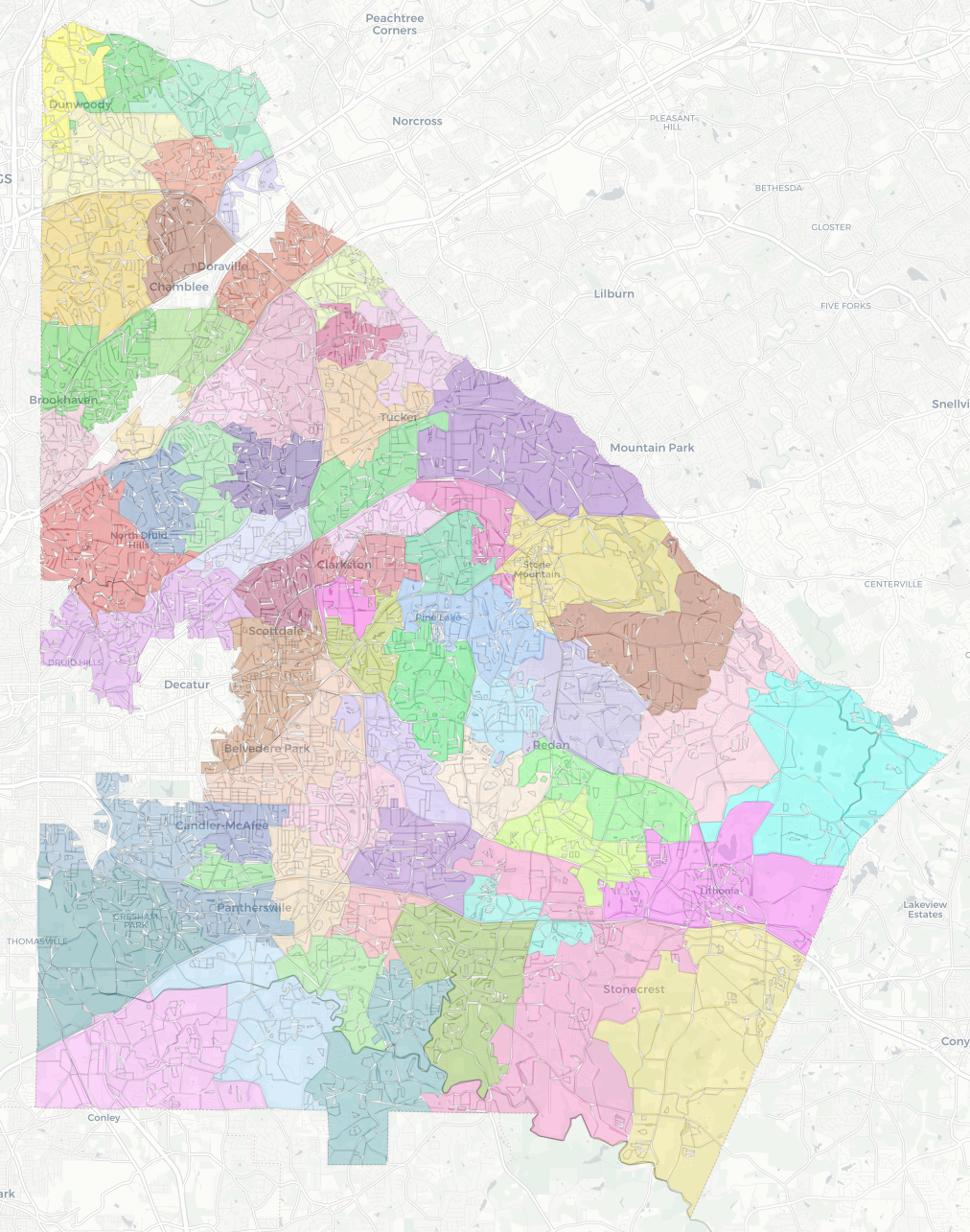}
    \end{subfigure}
    \hspace{1em}
    \begin{subfigure}{0.35\linewidth}
        \includegraphics[width=\linewidth]{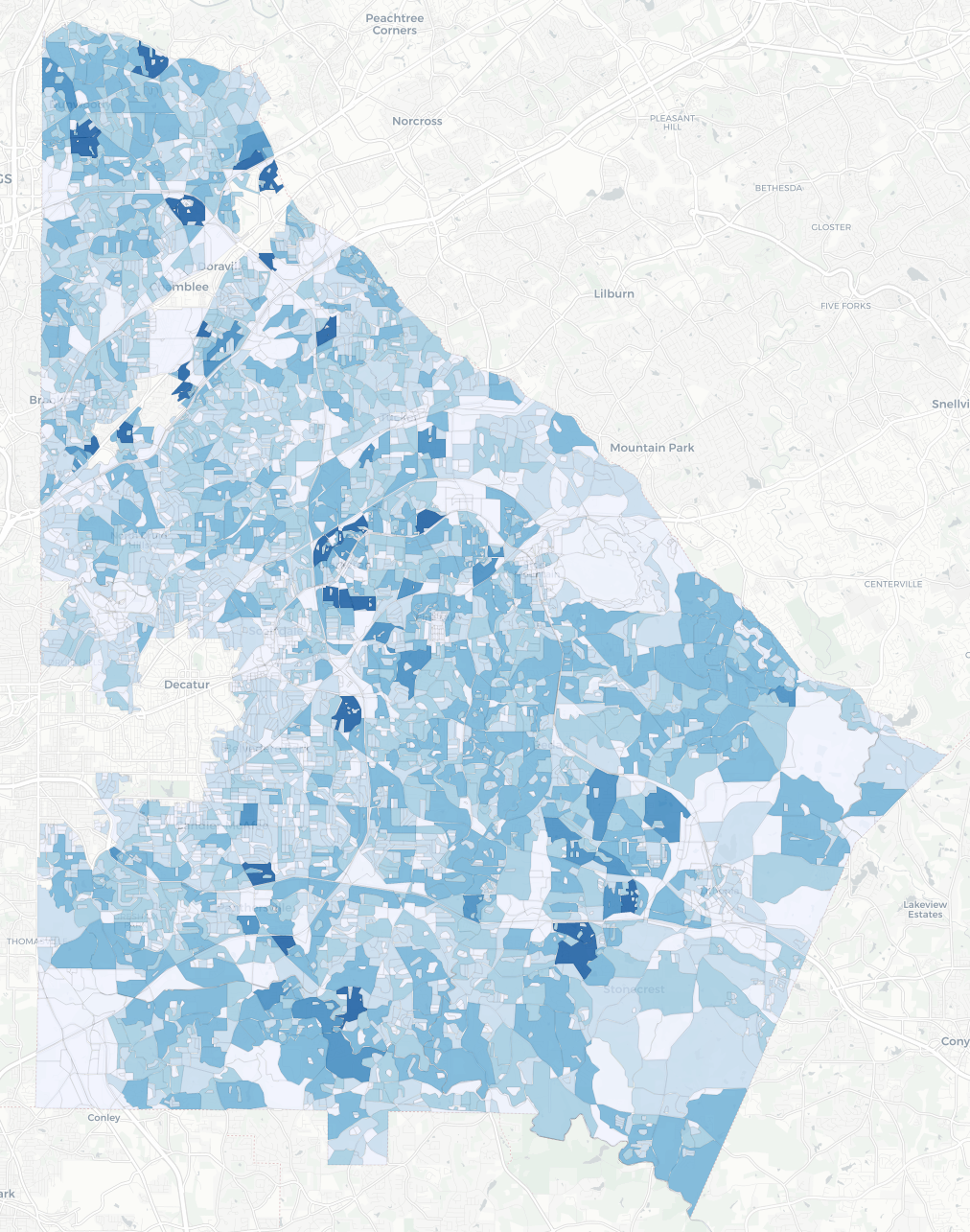}
    \end{subfigure}

    \vspace{.5em}
    \begin{subfigure}{0.35\linewidth}
        \includegraphics[width=\linewidth]{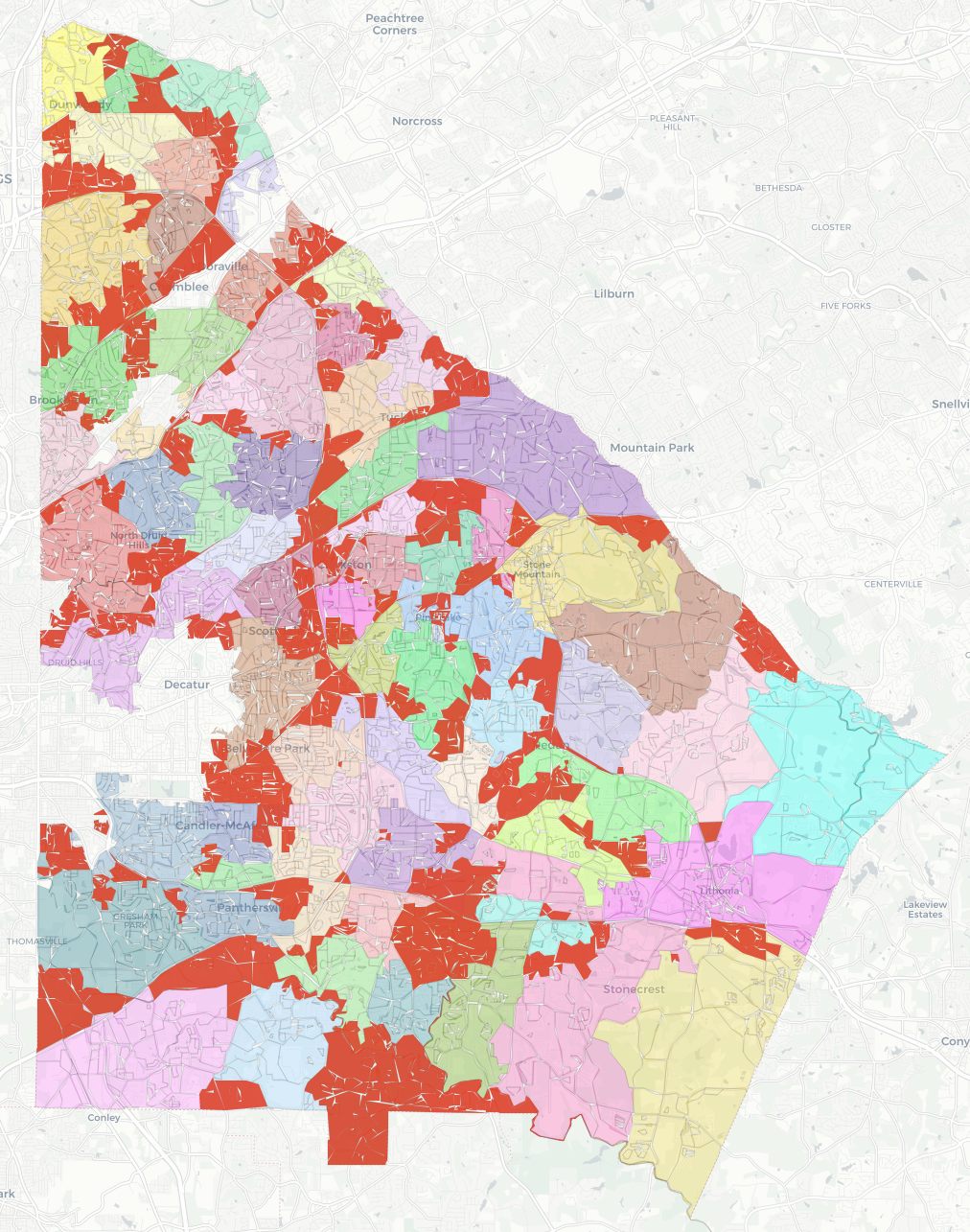}
    \end{subfigure}
    \hspace{1em}
    \begin{subfigure}{0.35\linewidth}
        \includegraphics[width=\linewidth]{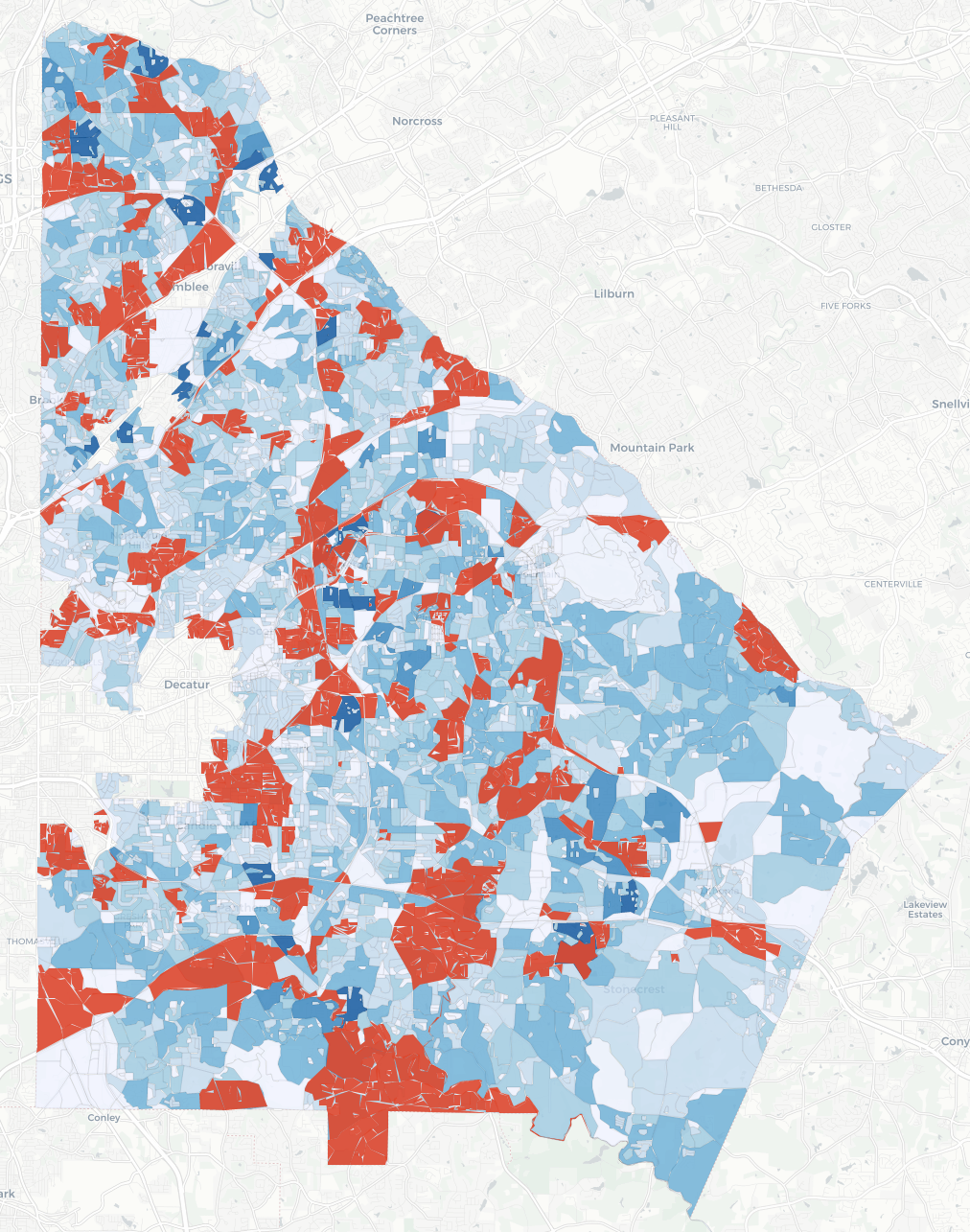}
    \end{subfigure}
    \vspace{.5em}

    \begin{subfigure}{0.35\linewidth}
        \includegraphics[width=\linewidth]{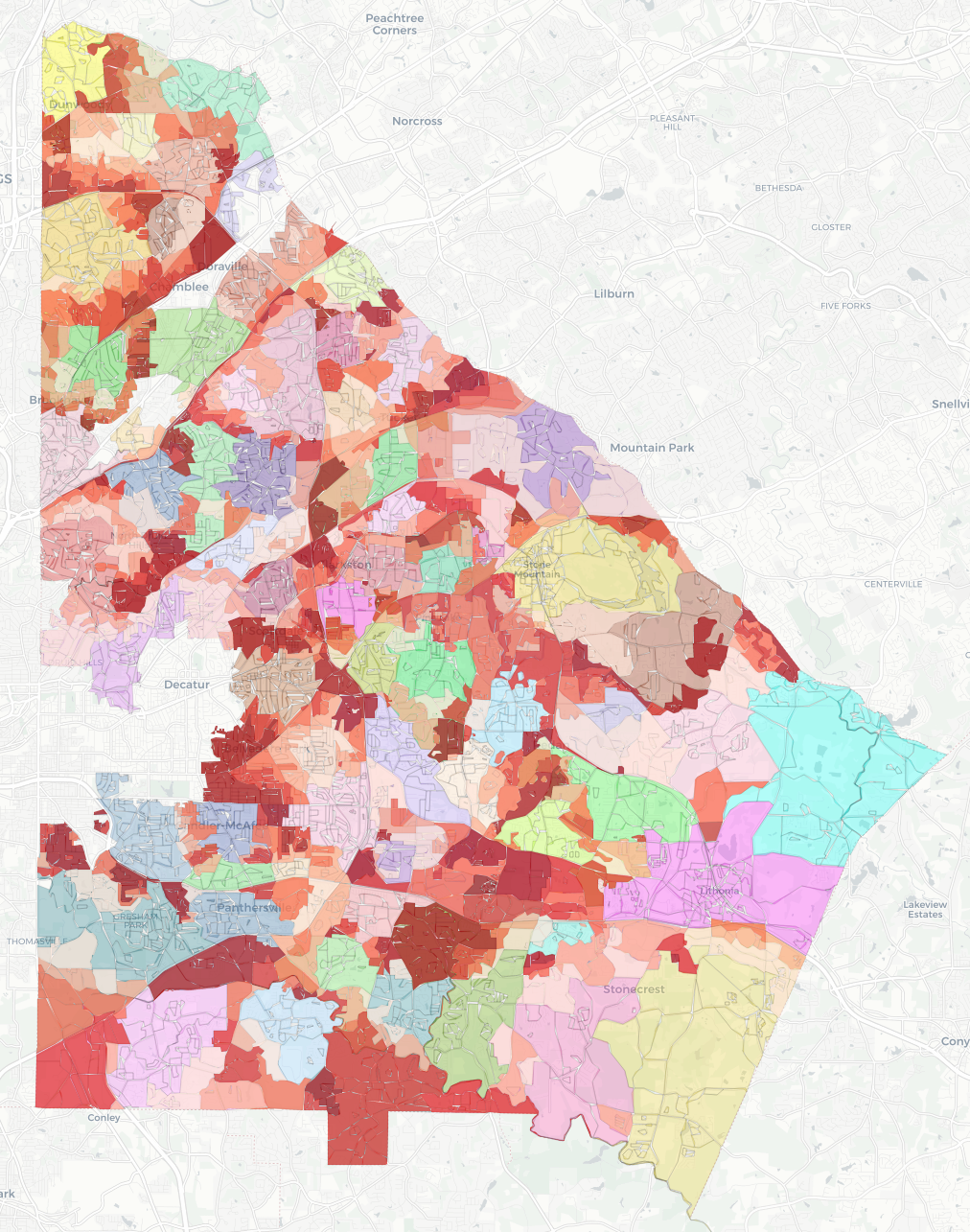}
    \end{subfigure}
    \hspace{1em}
    \begin{subfigure}{0.35\linewidth}
        \includegraphics[width=\linewidth]{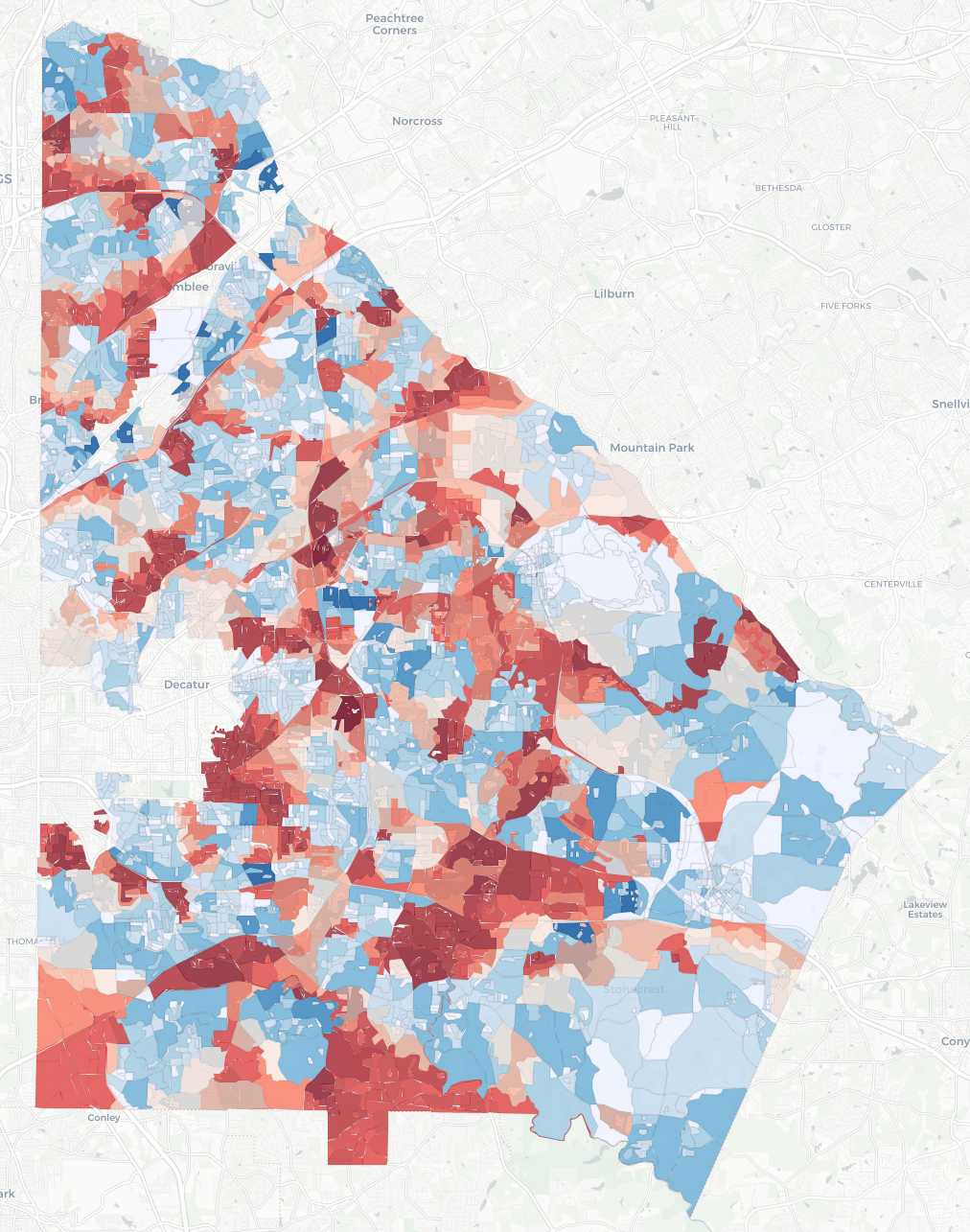}
    \end{subfigure}
    
    \caption{The two columns above present the layers of the current school
    assignment $\currassign$ and the population sizes $\bm{N}_{T,\cdot}$
    respectively for DeKalb County Schools.
    The choropleth maps in the second column use color shades to represent the 
    population size of each Census block, with darker shades of blue indicating higher numbers of students residing in the block. Census blocks rezoned
    by the non-private are highlighted in red for the figures 
    in the second row. Likewise, the figures in the last row visualize the 
    probability of Census blocks being rezoned by the private school assignment
    with $\epsilon=2$, where darker shades of red indicate a higher likelihood of rezoning over 200 independent simulations.}
    \label{fig:sixplots_dekalb}
\end{figure}

The purpose of this section is to carry out a more detailed, Census block-level analyses through case studies of two large, metropolitan, adjacent districts in Atlanta: DeKalb County and Atlanta Public Schools, represented by Figures \ref{fig:sixplots_dekalb} and \ref{fig:sixplots} (deferred to the Appendix), respectively.  The figures compare the assignment of Census blocks to schools under status quo (top row), non-private assignment (middle), and private assignment (bottom row).

We observe that, for both of sets of alternative school attendance boundaries, 
most of the rezonings take place near
the boundaries associated with the current school assignment, which 
is likely due to the travel time constraints imposed on the model.
Moreover, it is noteworthy that the non-private and private school assignments 
are often unlikely to re-assign highly-populated Census blocks.
This phenomenon can be explained by
the fact that the rezoning of a block with a high population might be more likely to violate the school size constraint (presented in Equation \eqref{eq:school_size}), or perhaps adversely affect the integration objective (if they have high concentrations of students belonging to a particular demographic group).
The ``average-case" private school assignment exhibits similar behavior
as the non-private school assignment at the level of Census blocks, though still with differences, reflecting comparisons between the non-private and private assignments across the various outcome measures discussed above. 

\section{Discussion}
\label{sec:discussion}
Overall, our results demonstrate that, as privacy protection strengthens,
private school assignment re-assigns fewer students and 
yields less diverse attendance boundaries. 
Furthermore, only a small amount of the 
difference between private and non-private school assignment's impact on changes in diversity can be
explained by district-level demographics and baseline segregation rates, suggesting the nuances of each district's current
boundaries and population distribution are likely to influence how much the introduction of differential privacy 
will affect the outcomes of redistricting to foster more diverse and integrated schools.  These findings point to a privacy-diversity trade-off local
educational policymakers may face in forthcoming years, particularly as computational
methods increasingly play a role in attendance boundary redrawing.  Interestingly, adding
differential privacy does not
exert large disparate impacts on travel time and percents of the students rezoned 
across different groups. 

Before concluding, we note several limitations with our work.  For one, our focus on ``average-case" analyses across
multiple districts may obfuscate district-level nuances vis-á-vis how much differential privacy is likely to affect policymaking in practice.  Focusing only on modeling a limited number of privacy budgets may have a similar impact.  We also estimate student counts by Census block (instead of injecting privacy-preserving noise on ground truth student counts, which would require closer collaborations and data agreements with school districts), which could affect the results presented here.  These limitations undescore the need for researchers to work closely with school districts in order to more thoroughly understand how the introduction of differential privacy might impact their diversity-promoting redistricting policies.

Looking ahead, we believe our results add to the growing empirical literature on how the introduction of differential privacy might affect decision-making in high-stakes policymaking settings.  School districts often redraw attendance boundaries in order to mitigate school utilization imbalances; respond to changing demographics and populations; accommodate new schools; and address several other topics.  Diversity is sometimes, though not always, a consideration in these settings.  With demographic segregation continuing to plague many districts across the US~\cite{gao2022segregation}---and issues of student admissions and diversity taking center stage in the US~\cite{affirmative}---there is a growing need to consider boundary-based student policy assignment changes in K12 settings in order to foster more diverse and integrated schools.  Fortunately, many districts are interested in exploring such changes~\cite{gillani2023las}; still, they may also lack the in-house expertise to do so, or budgets and other resources to engage with outside consultants (who themselves may lack the expertise to model and explore a variety of integration-promoting boundary changes).  This offers a unique opportunity for researchers and technologists to develop systems that support districts in these endeavors.  Making this scalable to and inclusive of the thousands of districts across the US will require the development of analytical platforms and tools that equip districts to explore their own potential boundary changes and associated downstream impacts on family and district-level outcomes they might care about.  Of course, garnering districts' trust to use these tools with potentially sensitive student data will be of paramount importance.  Imposing differential privacy can serve as a powerful avenue for garnering such trust, yet as we've shown, may also affect the extent to which districts can achieve diversity-promoting boundaries.  We hope our study offers preliminary insights into what these effects of differential privacy might be and how they might vary across districts, and inspires future work that seeks to carefully balance privacy, diversity, and equity in education policymaking settings.

% \begin{acks}
% \end{acks}

\bibliographystyle{ACM-Reference-Format}
\bibliography{cite}
\newpage

\appendix

\section{Supplementary figures}

\begin{figure}[h!]
    \centering
    \includegraphics[width=.7\linewidth]{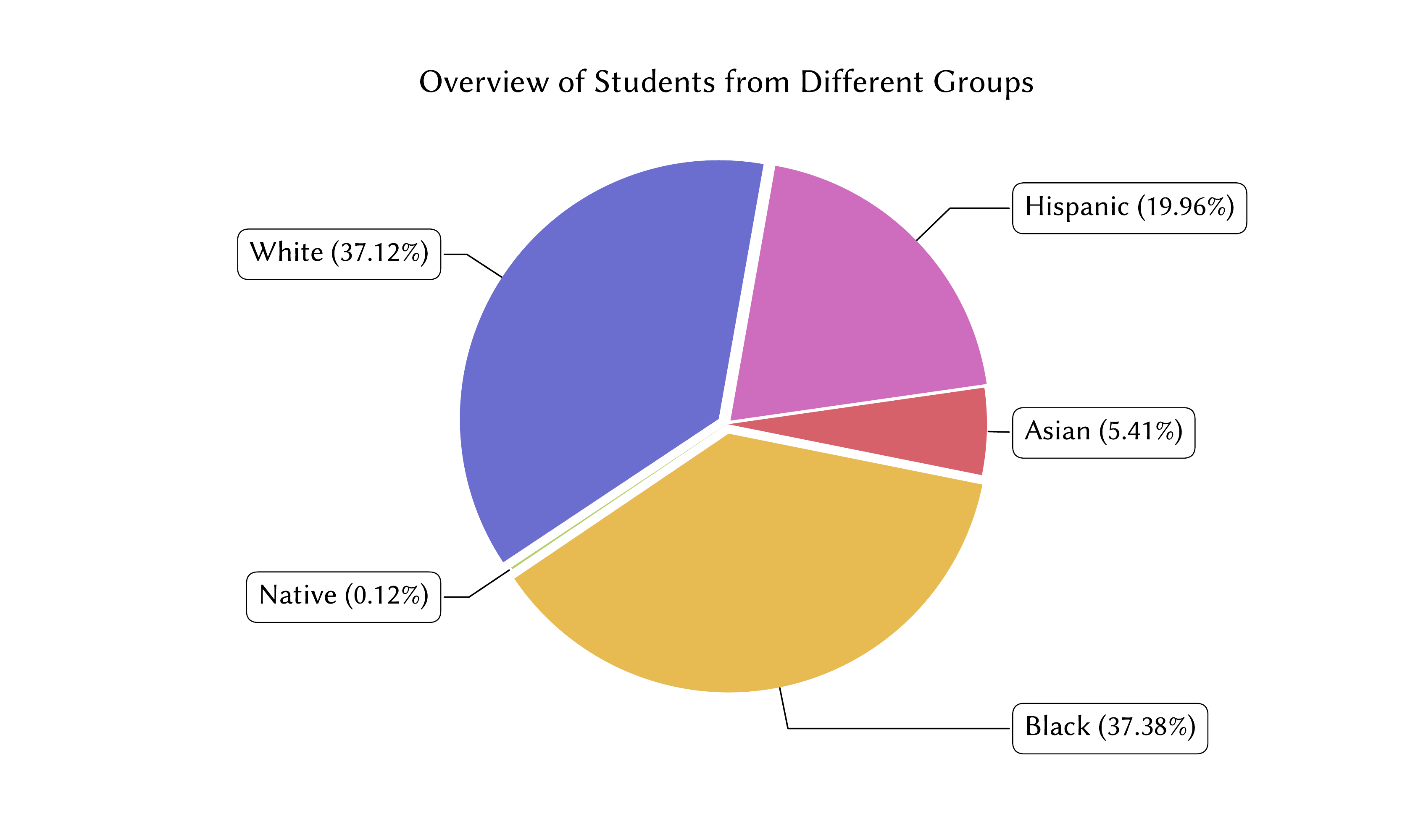}
    \caption{Demographic composition of students attending schools from the 67 districts in Georgia.}
    \label{fig:demo_info}
\end{figure}

\newpage
\begin{center}
\begin{longtable}[!h]{|r|r|r|r|r|r|r|}
\caption{Overview of the 67 School Districts in Georgia, along with the 
dissimilarity indices under different school assignments. The
last column contains the empirical mean estimations of the dissimilarity
indices associated with the private school assignments ($\epsilon=2$) over 200 independent repetitions.} \label{tab:long_dissim_idx_info} \\

\hline \multicolumn{1}{|r|}{\textbf{District ID}} & \multicolumn{1}{r|}{\textbf{School Districts}} 
& \multicolumn{1}{r|}{\textbf{\# Blocks}} & \multicolumn{1}{r|}{\textbf{\# Elem. Schools}}& 
\multicolumn{1}{r|}{$\di{\currassign}$}& \multicolumn{1}{r|}{$\di{\assign{\bm{N}}}$} 
& \multicolumn{1}{r|}{$\empe{\di{\assign{\tilde{\bm{N}}}}}$}\\ \hline 
\endfirsthead

\multicolumn{7}{c}%
{{\bfseries \tablename\ \thetable{} -- Continued from previous page}} \\
\hline \multicolumn{1}{|r|}{\textbf{District ID}} & \multicolumn{1}{r|}{\textbf{School Districts}} 
& \multicolumn{1}{r|}{\textbf{\# Blocks}} & \multicolumn{1}{r|}{\textbf{\# Elem. Schools}}& 
\multicolumn{1}{r|}{$\di{\currassign}$}& \multicolumn{1}{r|}{$\di{\assign{\bm{N}}}$} 
& \multicolumn{1}{r|}{$\ee{\di{\assign{\tilde{\bm{N}}}}}$}\\ \hline 
\endhead

\hline \multicolumn{7}{|r|}{{Continued on next page}} \\ \hline
\endfoot

\hline \hline
\endlastfoot

1300001 & Troup County & 1309  & 8     & 0.3814  & 0.3438  & 0.3593  \\
1300090 & Atkinson County & 545   & 2     & 0.0404  & 0.0228  & 0.0325  \\
1300120 & Atlanta Public Schools & 5939  & 44    & 0.7472  & 0.6813  & 0.7179  \\
1300290 & Barrow County & 1088  & 8     & 0.1031  & 0.0626  & 0.0757  \\
1300330 & Bartow County & 1862  & 12    & 0.2598  & 0.1698  & 0.1949  \\
1300420 & Bibb County & 3436  & 19    & 0.5511  & 0.5277  & 0.5320  \\
1300630 & Bulloch County & 2499  & 9     & 0.4282  & 0.3371  & 0.3563  \\
1300690 & Butts County & 664   & 3     & 0.1309  & 0.1000  & 0.1288  \\
1300840 & Carroll County & 2081  & 12    & 0.1830  & 0.1125  & 0.1316  \\
1300930 & Catoosa County & 1308  & 8     & 0.2372  & 0.1780  & 0.2128  \\
1301020 & Savannah-Chatham County & 5650  & 24    & 0.4517  & 0.4316  & 0.4146  \\
1301080 & Chattooga County & 841   & 3     & 0.3202  & 0.1687  & 0.2512  \\
1301110 & Cherokee County & 2959  & 23    & 0.2897  & 0.1980  & 0.2193  \\
1301170 & Clarke County & 1080  & 8     & 0.5001  & 0.3551  & 0.3828  \\
1301230 & Clayton County & 3060  & 35    & 0.2856  & 0.2088  & 0.2625  \\
1301290 & Cobb County & 6706  & 64    & 0.4899  & 0.4452  & 0.4516  \\
1301350 & Coffee County & 1701  & 8     & 0.2715  & 0.1025  & 0.1515  \\
1301380 & Colquitt County & 1813  & 10    & 0.3509  & 0.2295  & 0.2721  \\
1301410 & Columbia County & 1606  & 18    & 0.2111  & 0.1684  & 0.1785  \\
1301500 & Coweta County & 2512  & 19    & 0.3068  & 0.2051  & 0.2484  \\
1301590 & Dade County & 707   & 2     & 0.0720  & 0.0313  & 0.0596  \\
1301620 & Dalton Public Schools & 688   & 6     & 0.4659  & 0.2840  & 0.3367  \\
1301650 & Dawson County & 779   & 4     & 0.2041  & 0.1474  & 0.1834  \\
1301740 & DeKalb County & 5916  & 66    & 0.7209  & 0.6631  & 0.6853  \\
1301830 & Dougherty County & 2435  & 11    & 0.4762  & 0.4359  & 0.4636  \\
1301860 & Douglas County & 1548  & 20    & 0.3823  & 0.3286  & 0.3342  \\
1301980 & Effingham County & 1363  & 8     & 0.1760  & 0.1205  & 0.1353  \\
1302130 & Fayette County & 1549  & 14    & 0.4015  & 0.3038  & 0.3140  \\
1302190 & Floyd County & 1724  & 7     & 0.3101  & 0.2813  & 0.2980  \\
1302220 & Forsyth County & 1574  & 21    & 0.3760  & 0.3134  & 0.3162  \\
1302280 & Fulton County & 6278  & 59    & 0.6133  & 0.5494  & 0.5591  \\
1302310 & Gainesville City & 744   & 5     & 0.4287  & 0.2961  & 0.3196  \\
1302400 & Glynn County & 2495  & 10    & 0.4032  & 0.3284  & 0.3767  \\
1302430 & Gordon County & 1204  & 6     & 0.1252  & 0.0968  & 0.1157  \\
1302460 & Grady County & 1309  & 5     & 0.4846  & 0.2978  & 0.4181  \\
1302520 & Griffin-Spalding County & 1515  & 11    & 0.4330  & 0.2930  & 0.3302  \\
1302550 & Gwinnett County & 6536  & 80    & 0.4280  & 0.3868  & 0.3982  \\
1302580 & Habersham County & 1477  & 8     & 0.3999  & 0.3465  & 0.3720  \\
1302610 & Hall County & 3055  & 19    & 0.5331  & 0.4091  & 0.4436  \\
1302700 & Harris County & 1019  & 4     & 0.1522  & 0.1075  & 0.1383  \\
1302730 & Hart County & 1214  & 3     & 0.2012  & 0.1473  & 0.1845  \\
1302790 & Heard County & 543   & 3     & 0.1530  & 0.0922  & 0.1430  \\
1302820 & Henry County & 3153  & 27    & 0.4671  & 0.3846  & 0.4029  \\
1302880 & Houston County & 3267  & 20    & 0.3307  & 0.2752  & 0.2904  \\
1302940 & Jackson County & 1550  & 6     & 0.1428  & 0.1128  & 0.1275  \\
1303150 & Jones County & 692   & 4     & 0.2469  & 0.1363  & 0.1762  \\
1303300 & Liberty County & 1189  & 7     & 0.2089  & 0.1718  & 0.1865  \\
1303390 & Lowndes County & 1779  & 7     & 0.1873  & 0.1303  & 0.1578  \\
1303480 & Madison County & 1140  & 5     & 0.2791  & 0.2348  & 0.2550  \\
1303510 & Marietta City & 972   & 7     & 0.6007  & 0.3581  & 0.4156  \\
1303720 & Monroe County & 994   & 3     & 0.1051  & 0.0517  & 0.0886  \\
1303840 & Murray County & 788   & 6     & 0.2339  & 0.0947  & 0.1250  \\
1303870 & Muscogee County & 2963  & 30    & 0.5389  & 0.4897  & 0.5039  \\
1303930 & Newton County & 1930  & 13    & 0.4578  & 0.3910  & 0.4137  \\
1303960 & Oconee County & 750   & 5     & 0.2040  & 0.1331  & 0.1827  \\
1304020 & Paulding County & 1508  & 19    & 0.2518  & 0.2167  & 0.2215  \\
1304050 & Peach County & 851   & 3     & 0.4192  & 0.3488  & 0.3910  \\
1304140 & Pierce County & 1103  & 3     & 0.2146  & 0.1277  & 0.1801  \\
1304200 & Polk County & 1524  & 6     & 0.2483  & 0.2180  & 0.2348  \\
1304380 & Richmond County & 3525  & 27    & 0.5369  & 0.4748  & 0.5032  \\
1304410 & Rockdale County & 1231  & 11    & 0.3336  & 0.2874  & 0.2983  \\
1304440 & Rome City & 807   & 5     & 0.4082  & 0.1465  & 0.1824  \\
1304980 & Tift County & 1874  & 8     & 0.1256  & 0.1109  & 0.1226  \\
1305310 & Valdosta City & 1131  & 5     & 0.4077  & 0.2133  & 0.2827  \\
1305370 & Walker County & 1923  & 10    & 0.1687  & 0.1443  & 0.1642  \\
1305390 & Walton County & 1666  & 9     & 0.3130  & 0.1857  & 0.2267  \\
1305700 & Whitfield County & 1683  & 13    & 0.4025  & 0.3420  & 0.3538  \\
\end{longtable}
\end{center}

\begin{figure}
    \centering
    \begin{subfigure}{0.4\linewidth}
        \includegraphics[width=\linewidth]{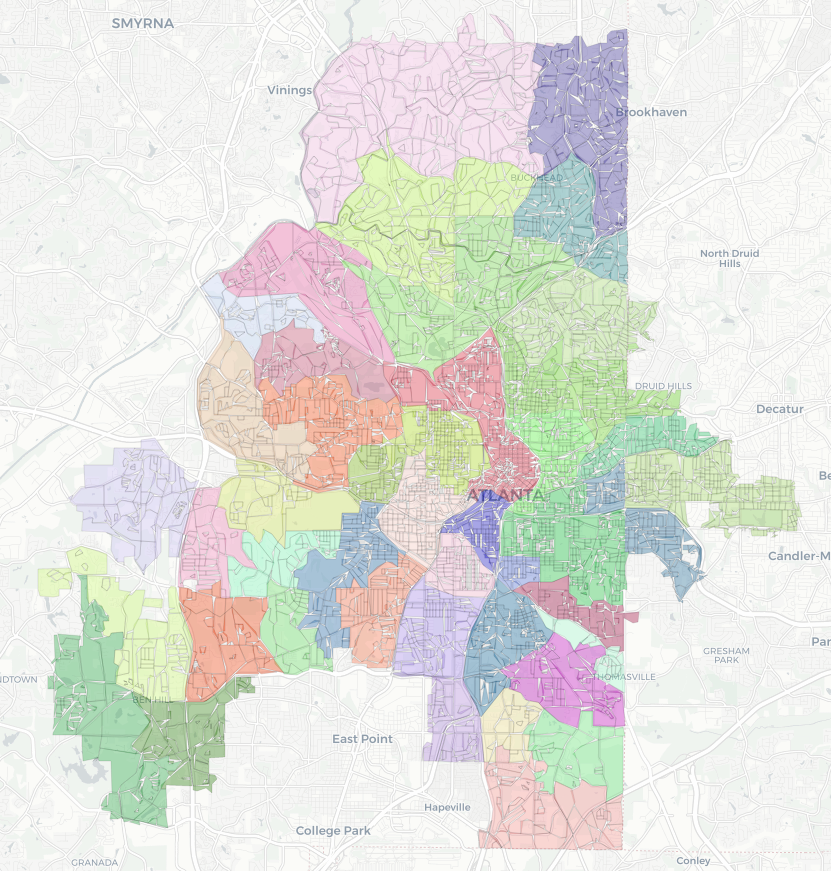}
        \label{fig:plot1}
    \end{subfigure}
    \hspace{1em}
    \begin{subfigure}{0.4\linewidth}
        \includegraphics[width=\linewidth]{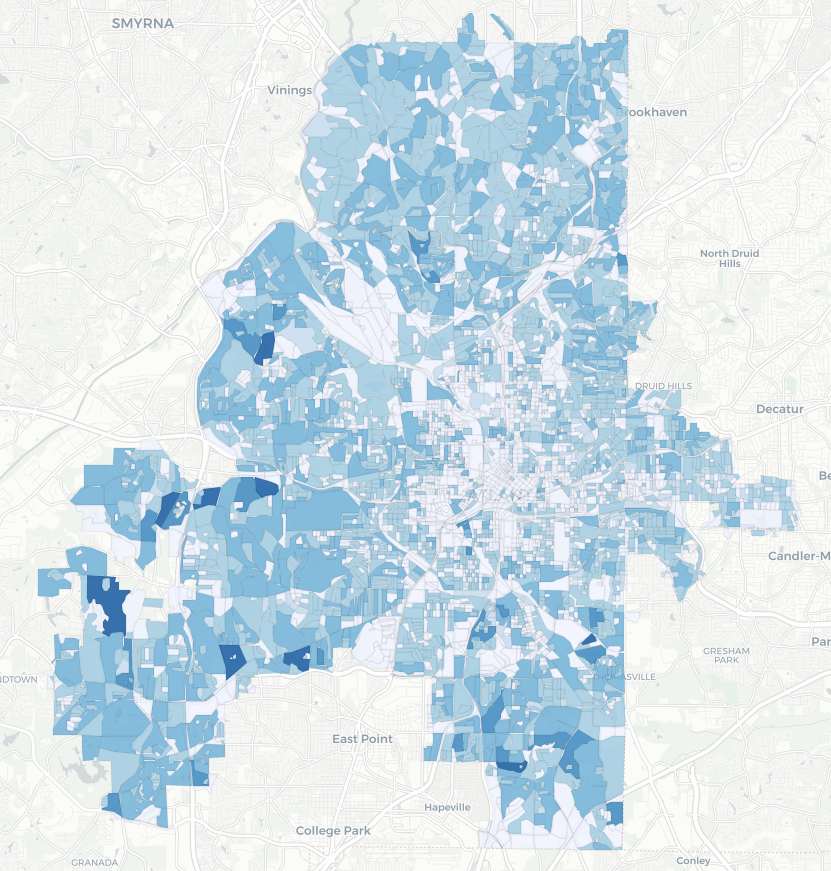}
        \label{fig:plot2}
    \end{subfigure}

    \begin{subfigure}{0.4\linewidth}
        \includegraphics[width=\linewidth]{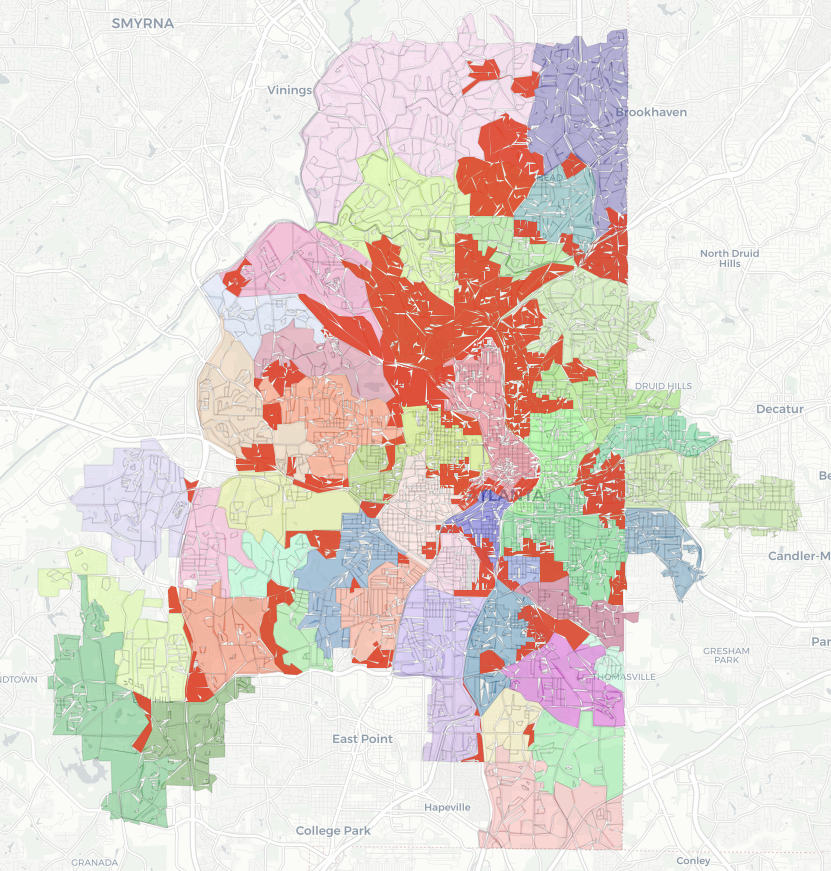}
        \label{fig:plot3}
    \end{subfigure}
    \hspace{1em}
    \begin{subfigure}{0.4\linewidth}
        \includegraphics[width=\linewidth]{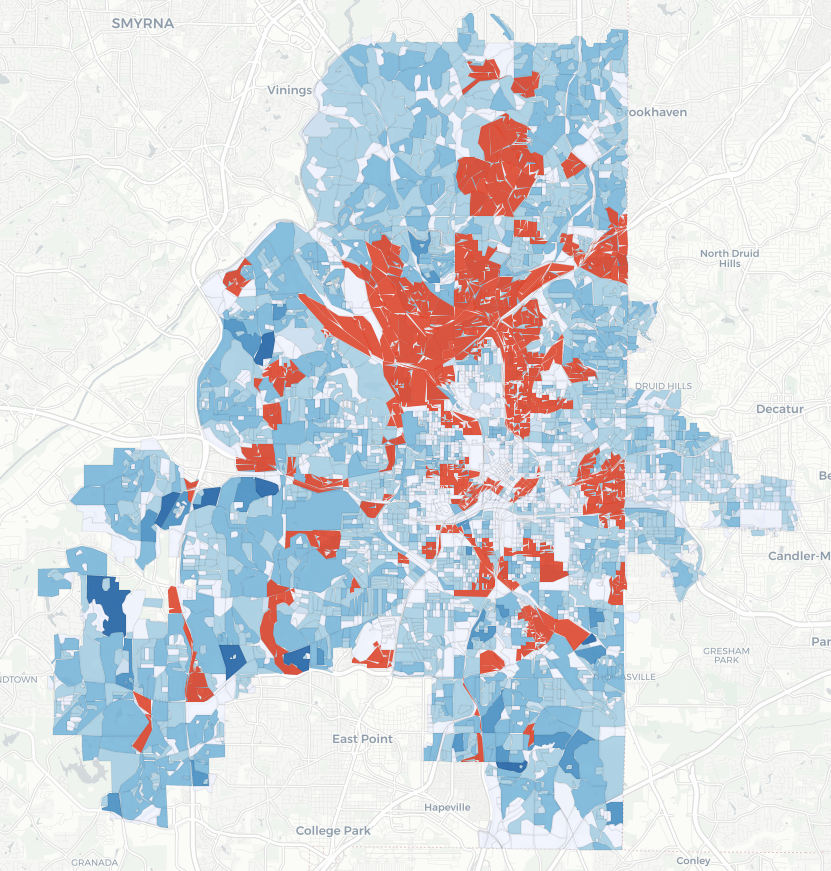}
        \label{fig:plot4}
    \end{subfigure}

    \begin{subfigure}{0.4\linewidth}
        \includegraphics[width=\linewidth]{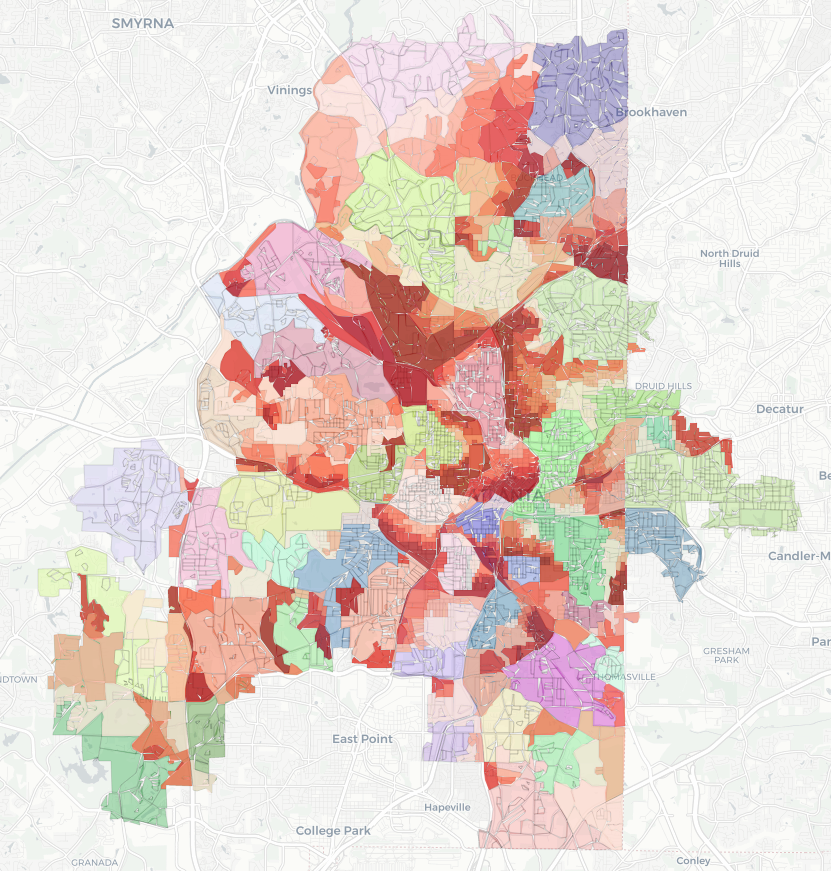}
        \label{fig:plot5}
    \end{subfigure}
    \hspace{1em}
    \begin{subfigure}{0.4\linewidth}
        \includegraphics[width=\linewidth]{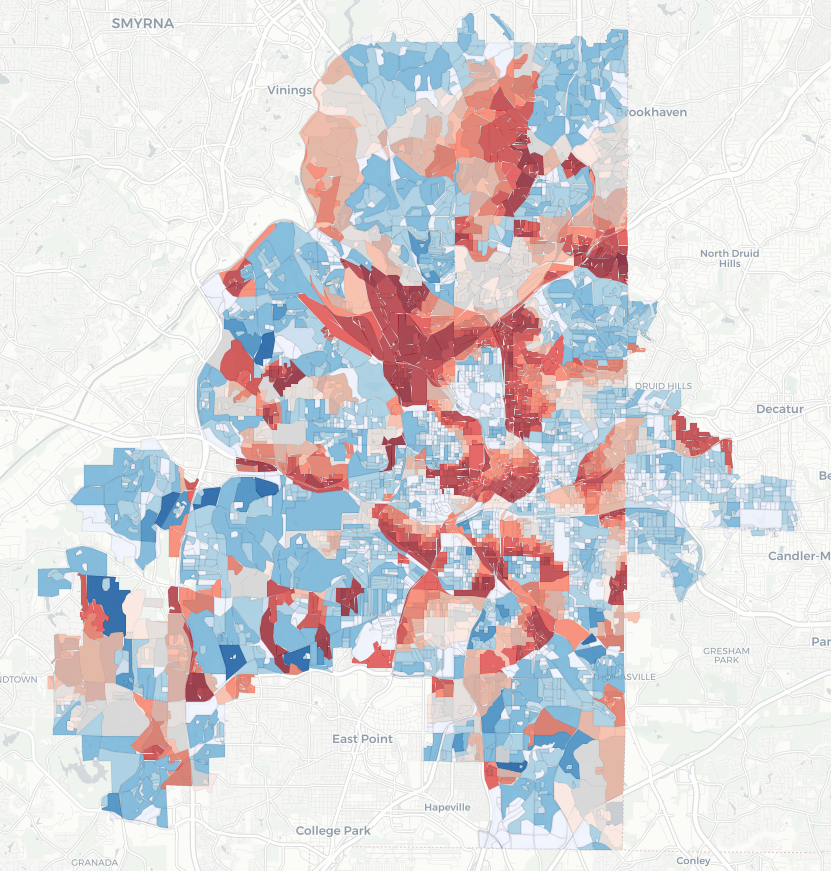}
        \label{fig:plot6}
    \end{subfigure}
    
    \caption{The two columns above present the layers of the current school
    assignment $\currassign$ and the population sizes $\bm{N}_{T,\cdot}$
    respectively for the school district, Atlanta Public Schools.
    The choropleth maps in the second column use color shades to represent the 
    population size of each Census block, with darker shades of blue indicating higher numbers of students residing in the block. Census blocks that are rezoned by the non-private are highlighted in red for the figures 
    in the second row. Likewise, the figures in the last row visualize the 
    probability of Census blocks being rezoned by the private school assignment
    with $\epsilon=2$, where darker shades of red indicate a higher likelihood of rezoning over 200 independent simulations.}
    \label{fig:sixplots}
\end{figure}

\section{Extension to socioeconomic (SES) integration}
Beyond racial and ethnic segregation, this section aims to explore 
the adoption of the optimization approach discussed in Section \ref{sec:problem}
to facilitate socioeconomic (SES) integration and investigate how differential privacy
affects the school attendance boundaries optimized for bringing together 
students of different socioeconomic statuses.

SES is a nebulous and multidimensional measure.  Districts will often use Free/Reduced-Price Lunch (FRL) eligibility as a proxy, yet FRL suffers from a number of limitations, including its often weak association with household economic resources~\cite{harwell2010frl}.  Here, we compute an index measure of SES used by several school districts across the US~\cite{quick2016cps}.  The measurecombines several Census block-group level demographic variables into a composite value.  These variables include: the percentage of dual parent households, percentage of households with a bachelor's degree, percentage of households speaking non-English languages, percentage of homes that are owner-occupied, and median family income.  We source these variables from the 2017-2021 American Community Survey (ACS) and standardize (z-score) each of these measures independently at the block group level.  The z-scores for each variable are averaged together, and a global z-score is computed over this average to arrive at an SES score for each block group.  Because our optimization model operates over blocks (which are finer-grained than block groups), we make the simplifying assumption of applying the block-group level z-scores to each of the block group's constituent blocks.  Blocks with z-scores falling above average are classified as high-SES; those falling below average are classified as low-SES.  An additional limitation of this SES measure (which districts themselves face when making policy decisions based on Census or ACS-derived data) is that it is not available at the student level; only at a coarser geographic level.   

Similar to the method presented in the main text, we apply differential privacy by injecting noise on the number of students per Census block belonging to each SES group (analogous to doing so on the number of students belonging to each racial/ethnic group).  Future work may involve exploring the impacts of injecting noise on variables upstream of these block-level counts per SES group---for example, on the individual variables that combine to produce the index SES measure.  

\begin{figure}
    \centering
    \includegraphics[width=.7\linewidth]{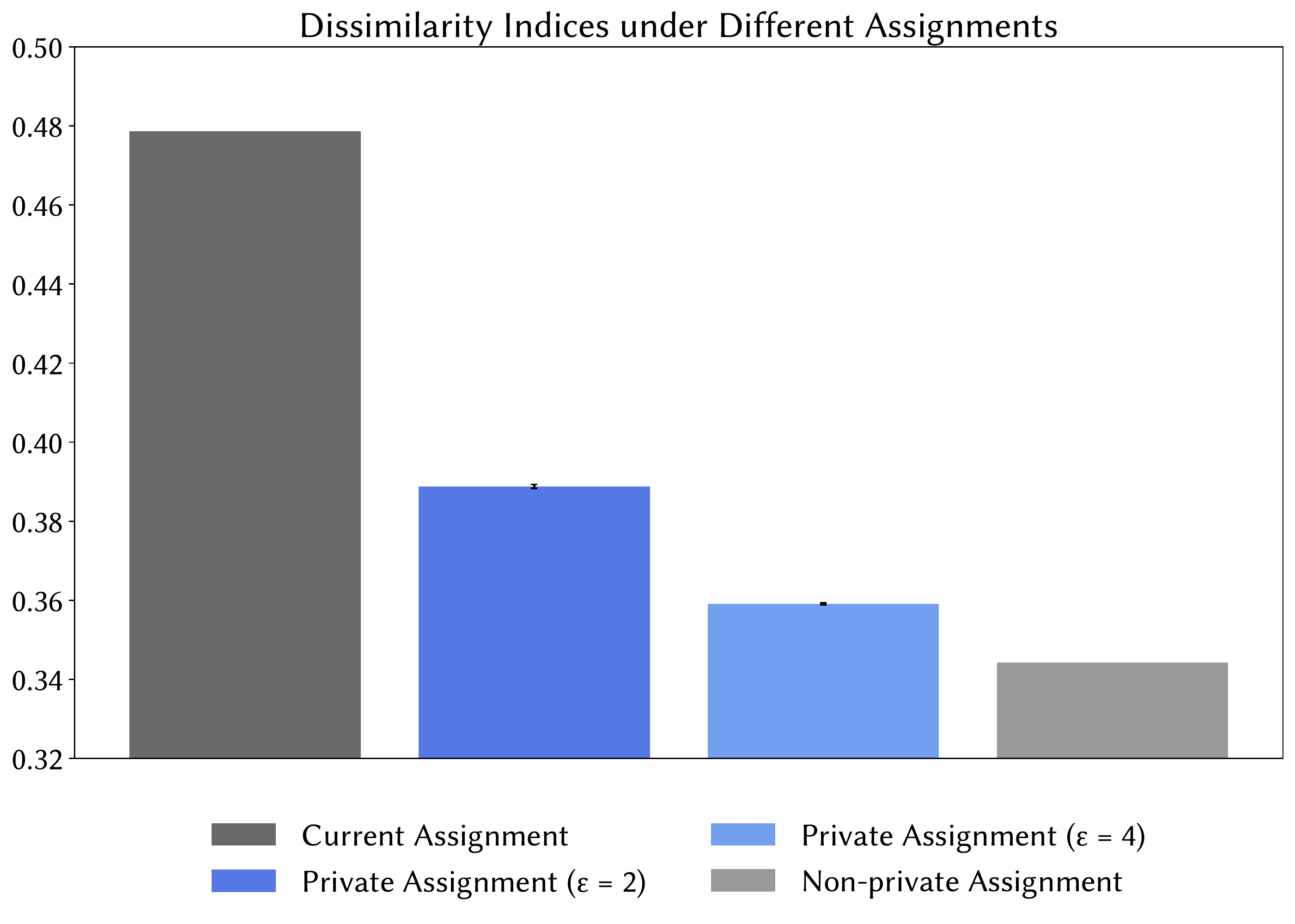}
    \caption{Average low-SES/high-SES 
    dissimilarity indices associated with current, non-private, and private 
    school assignments
    with  privacy budgets $\epsilon=2$ and $4$. Error bars depict 95\% bootstrapped
    confidence intervals for results across independent rezonings with random noise added.  The bars represent averages across districts; for a sense of variability by district, please see the extended Table~\ref{tab:ses_long_dissim_idx_info}.}
    \label{fig:ses_dissim}
\end{figure}

\begin{figure}
    \centering
    \includegraphics[width=.7\linewidth]{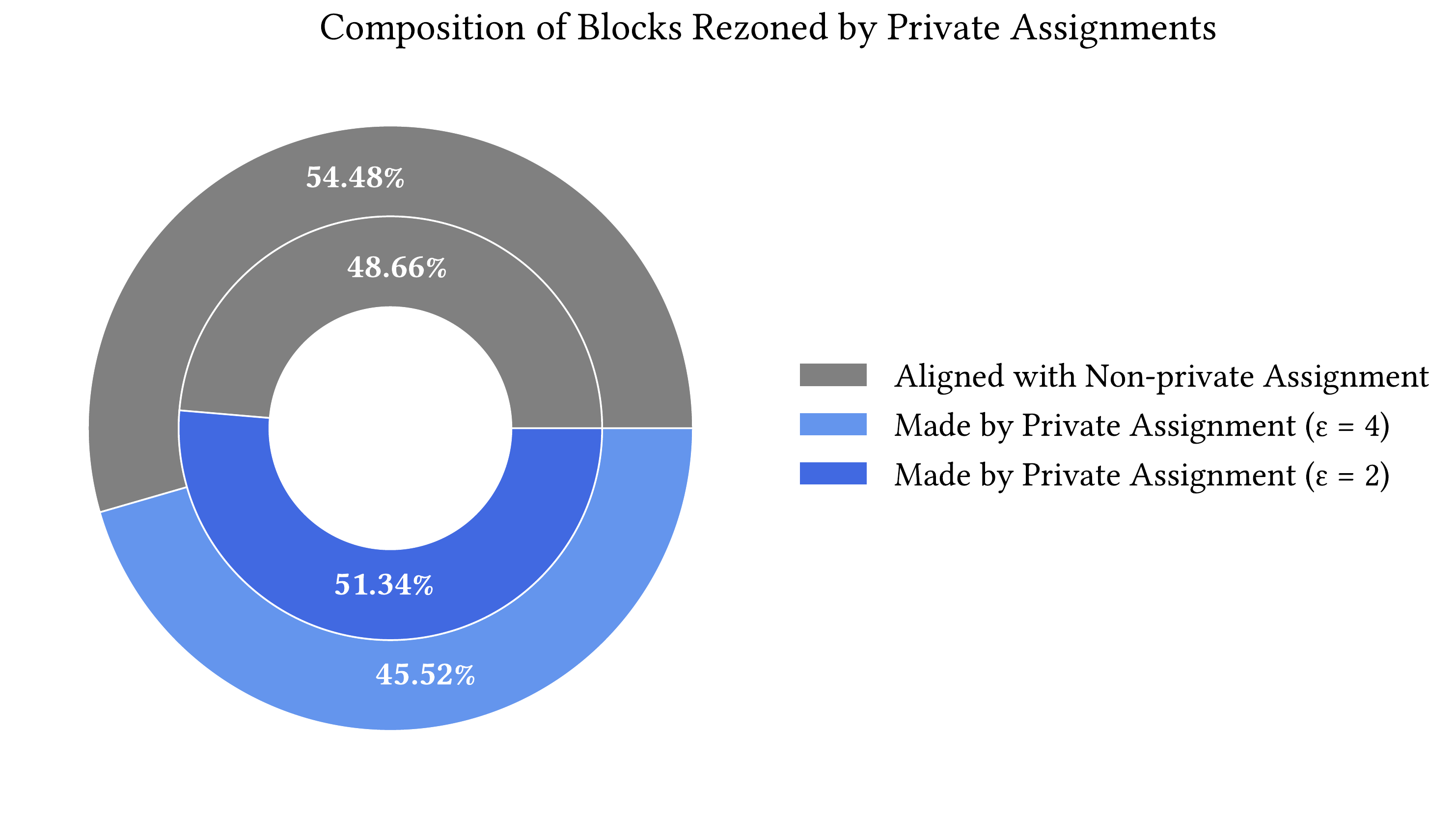}
    \caption{Composition of Census blocks rezoned by private 
        school assignments.
        The inner and outer circles correspond to the private
        school assignments with $\epsilon=2$ and $4$, respectively.}
    \label{fig:block_comp_ses}
\end{figure}

\begin{figure}
    \centering
    \includegraphics[width=.7\linewidth]{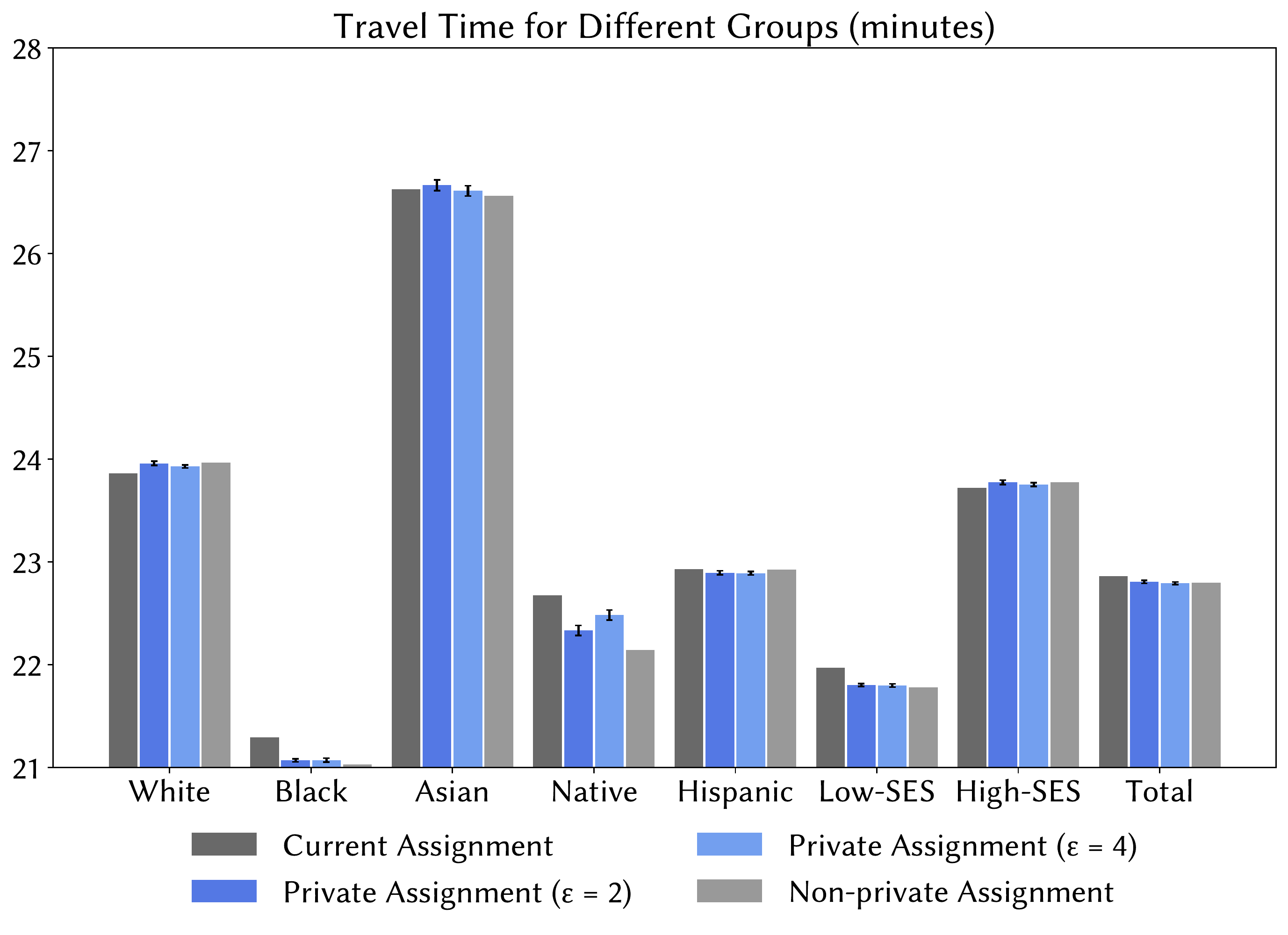}
    \caption{Travel time for different groups. Error bars depict 95\% bootstrapped confidence
    intervals across across independent privacy runs.}
    \label{fig:ses_travel}
\end{figure}

\begin{figure}
    \centering
    \includegraphics[width=.7\linewidth]{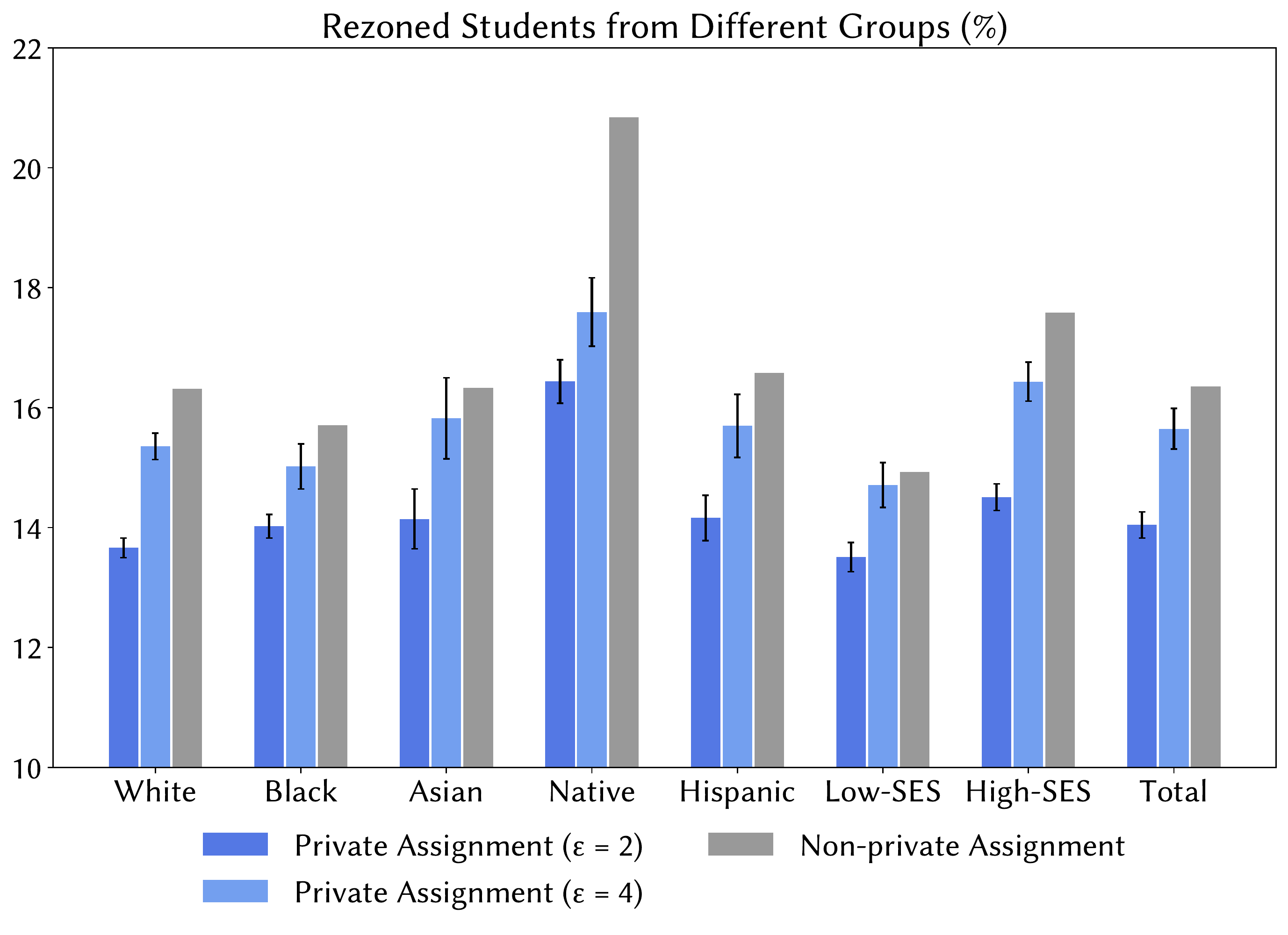}
    \caption{Percentage of the students rezoned across different groups. 
    Error bars depict 95\% bootstrapped confidence intervals across independent privacy runs.}
    \label{fig:ses_rezoned}
\end{figure}

In general, we find that the results produced from the SES simulations are consistent with those produced by the ones for race/ethnicity: stronger privacy guarantees might impede SES integration efforts (perhaps resulting from fewer students being rezoned), with minimal impacts on travel times.  Figures~\ref{fig:ses_dissim}-\ref{fig:ses_rezoned} and Table~\ref{tab:ses_long_dissim_idx_info} describe these results in greater detail, mirroring those presented in the main text.

% \begin{figure}
%     \centering
%     \includegraphics[width=.9\linewidth]{img/rezoned_blk_ses.png}
%     \caption{Number of Average Rezoned blocks.}
%     \label{fig:ses_blocks_rezoned}
% \end{figure}

\begin{center}
\begin{longtable}[!h]{|r|r|r|r|r|r|r|}
\caption{Overview of the 67 School Districts in Georgia, along with the 
dissimilarity indices under different school assignments. The last two
columns report the dissimilarity indices associated with the school
assignments optimized for socioeconomic integration.} \label{tab:ses_long_dissim_idx_info} \\

\hline \multicolumn{1}{|r|}{\textbf{District ID}} & \multicolumn{1}{r|}{\textbf{School Districts}} 
& \multicolumn{1}{r|}{\textbf{\# Blocks}} & \multicolumn{1}{r|}{\textbf{\# Elem. Schools}}& 
\multicolumn{1}{r|}{$\di{\currassign}$}& \multicolumn{1}{r|}{$\di{\assign{\bm{N}}}$} 
& \multicolumn{1}{r|}{$\empe{\di{\assign{\tilde{\bm{N}}}}}$}\\ \hline 
\endfirsthead

\multicolumn{7}{c}%
{{\bfseries \tablename\ \thetable{} -- Continued from previous page}} \\
\hline \multicolumn{1}{|r|}{\textbf{District ID}} & \multicolumn{1}{r|}{\textbf{School Districts}} 
& \multicolumn{1}{r|}{\textbf{\# Blocks}} & \multicolumn{1}{r|}{\textbf{\# Elem. Schools}}& 
\multicolumn{1}{r|}{$\di{\currassign}$}& \multicolumn{1}{r|}{$\di{\assign{\bm{N}}}$} 
& \multicolumn{1}{r|}{$\ee{\di{\assign{\tilde{\bm{N}}}}}$}\\ \hline 
\endhead

\hline \multicolumn{7}{|r|}{{Continued on next page}} \\ \hline
\endfoot

\hline \hline
\endlastfoot

        1300001 & Troup County & 1309 & 8 & 0.6550 & 0.6027 & 0.6333 \\ 
        1300090 & Atkinson County & 545 & 2 & 0.4188 & 0.3876 & 0.4215 \\ 
        1300120 & Atlanta Public Schools & 5939 & 44 & 0.6882 & 0.6354 & 0.6557 \\ 
        1300290 & Barrow County & 1088 & 8 & 0.3966 & 0.2704 & 0.2927 \\ 
        1300330 & Bartow County & 1862 & 12 & 0.2557 & 0.1301 & 0.1629 \\ 
        1300420 & Bibb County & 3436 & 19 & 0.7467 & 0.6693 & 0.6821 \\ 
        1300630 & Bulloch County & 2499 & 9 & 0.4650 & 0.1286 & 0.2401 \\ 
        1300690 & Butts County & 664 & 3 & 0.4440 & 0.3056 & 0.3590 \\ 
        1300840 & Carroll County & 2081 & 12 & 0.3633 & 0.2101 & 0.2580 \\ 
        1300930 & Catoosa County & 1308 & 8 & 0.5671 & 0.5317 & 0.5632 \\ 
        1301020 & Savannah-Chatham County & 5650 & 24 & 0.6007 & 0.5718 & 0.5357 \\ 
        1301080 & Chattooga County & 841 & 3 & 0.2732 & 0.0329 & 0.1272 \\ 
        1301110 & Cherokee County & 2959 & 23 & 0.4662 & 0.3218 & 0.3611 \\ 
        1301170 & Clarke County & 1080 & 8 & 0.5486 & 0.3983 & 0.4370 \\ 
        1301230 & Clayton County & 3060 & 35 & 0.5176 & 0.3849 & 0.4141 \\ 
        1301290 & Cobb County & 6706 & 64 & 0.6658 & 0.5903 & 0.6193 \\ 
        1301350 & Coffee County & 1701 & 8 & 0.2748 & 0.0911 & 0.1376 \\ 
        1301380 & Colquitt County & 1813 & 10 & 0.5581 & 0.2978 & 0.3780 \\ 
        1301410 & Columbia County & 1606 & 18 & 0.5500 & 0.4713 & 0.4810 \\ 
        1301500 & Coweta County & 2512 & 19 & 0.5507 & 0.3521 & 0.4057 \\ 
        1301590 & Dade County & 707 & 2 & 0.0908 & 0.0009 & 0.0640 \\ 
        1301620 & Dalton Public Schools & 688 & 6 & 0.7054 & 0.5535 & 0.5812 \\ 
        1301650 & Dawson County & 779 & 4 & 0.2156 & 0.1481 & 0.2393 \\ 
        1301740 & DeKalb County & 5916 & 66 & 0.5722 & 0.5101 & 0.5325 \\ 
        1301830 & Dougherty County & 2435 & 11 & 0.5133 & 0.4501 & 0.5090 \\ 
        1301860 & Douglas County & 1548 & 20 & 0.5213 & 0.3828 & 0.4197 \\ 
        1301980 & Effingham County & 1363 & 8 & 0.4301 & 0.3608 & 0.3666 \\ 
        1302130 & Fayette County & 1549 & 14 & 0.4889 & 0.2661 & 0.3033 \\ 
        1302190 & Floyd County & 1724 & 7 & 0.5490 & 0.3487 & 0.4308 \\ 
        1302220 & Forsyth County & 1574 & 21 & 0.4978 & 0.3092 & 0.3287 \\ 
        1302280 & Fulton County & 6278 & 59 & 0.5710 & 0.4951 & 0.5131 \\ 
        1302310 & Gainesville City & 744 & 5 & 0.3972 & 0.1802 & 0.1844 \\ 
        1302400 & Glynn County & 2495 & 10 & 0.5880 & 0.4942 & 0.5360 \\ 
        1302430 & Gordon County & 1204 & 6 & 0.5733 & 0.4379 & 0.4566 \\ 
        1302460 & Grady County & 1309 & 5 & 0.3614 & 0.2094 & 0.3217 \\ 
        1302520 & Griffin-Spalding County & 1515 & 11 & 0.5679 & 0.3170 & 0.3943 \\ 
        1302550 & Gwinnett County & 6536 & 80 & 0.6223 & 0.5661 & 0.6065 \\ 
        1302580 & Habersham County & 1477 & 8 & 0.4447 & 0.3249 & 0.3438 \\ 
        1302610 & Hall County & 3055 & 19 & 0.7070 & 0.5295 & 0.5884 \\ 
        1302700 & Harris County & 1019 & 4 & 0.3599 & 0.2834 & 0.3301 \\ 
        1302730 & Hart County & 1214 & 3 & 0.1880 & 0.0269 & 0.1190 \\ 
        1302790 & Heard County & 543 & 3 & 0.8718 & 0.8235 & 0.8750 \\ 
        1302820 & Henry County & 3153 & 27 & 0.5625 & 0.3612 & 0.3918 \\ 
        1302880 & Houston County & 3267 & 20 & 0.5296 & 0.4434 & 0.4610 \\ 
        1302940 & Jackson County & 1550 & 6 & 0.6178 & 0.4955 & 0.5397 \\ 
        1303150 & Jones County & 692 & 4 & 0.2930 & 0.1100 & 0.1795 \\ 
        1303300 & Liberty County & 1189 & 7 & 0.5862 & 0.4264 & 0.4521 \\ 
        1303390 & Lowndes County & 1779 & 7 & 0.2946 & 0.2035 & 0.2435 \\ 
        1303480 & Madison County & 1140 & 5 & 0.6253 & 0.4248 & 0.5183 \\ 
        1303510 & Marietta City & 972 & 7 & 0.5551 & 0.3705 & 0.4442 \\ 
        1303720 & Monroe County & 994 & 3 & 0.4079 & 0.1540 & 0.3526 \\ 
        1303840 & Murray County & 788 & 6 & 0.3981 & 0.2648 & 0.2779 \\ 
        1303870 & Muscogee County & 2963 & 30 & 0.5764 & 0.4743 & 0.4814 \\ 
        1303930 & Newton County & 1930 & 13 & 0.4510 & 0.3032 & 0.3112 \\ 
        1303960 & Oconee County & 750 & 5 & 0.3422 & 0.2051 & 0.2578 \\ 
        1304020 & Paulding County & 1508 & 19 & 0.5755 & 0.4587 & 0.4699 \\ 
        1304050 & Peach County & 851 & 3 & 0.4691 & 0.3225 & 0.3975 \\ 
        1304140 & Pierce County & 1103 & 3 & 0.3128 & 0.2162 & 0.2470 \\ 
        1304200 & Polk County & 1524 & 6 & 0.2722 & 0.1615 & 0.1958 \\ 
        1304380 & Richmond County & 3525 & 27 & 0.5363 & 0.4025 & 0.4384 \\ 
        1304410 & Rockdale County & 1231 & 11 & 0.3833 & 0.2606 & 0.2712 \\ 
        1304440 & Rome City & 807 & 5 & 0.2453 & 0.1823 & 0.1982 \\ 
        1304980 & Tift County & 1874 & 8 & 0.2417 & 0.1179 & 0.1545 \\ 
        1305310 & Valdosta City & 1131 & 5 & 0.6572 & 0.3995 & 0.4707 \\ 
        1305370 & Walker County & 1923 & 10 & 0.5199 & 0.4066 & 0.4950 \\ 
        1305390 & Walton County & 1666 & 9 & 0.3276 & 0.1397 & 0.2198 \\ 
        1305700 & Whitfield County & 1683 & 13 & 0.4438 & 0.3605 & 0.3834 \\ 
\end{longtable}
\end{center}

\end{document}